\documentclass[a4paper, 11pt]{article}
\usepackage{fullpage}

\usepackage[utf8]{inputenc}
\usepackage[T1]{fontenc}

\usepackage{amsthm}
\usepackage{amssymb}
\usepackage{thmtools}
\usepackage{thm-restate}
\usepackage{latexsym}
\usepackage{MnSymbol}

\usepackage{subcaption}
\usepackage{comment}

\usepackage{tikz}
\usetikzlibrary{positioning}
\usetikzlibrary{decorations.pathmorphing}

\renewcommand{\S}{{\cal{S}}}
\newcommand{\C}{{\cal{C}}}
\newcommand{\Q}{{\cal{Q}}}
\newcommand{\R}{{\cal{R}}}
\newcommand{\T}{{\cal{T}}}
\renewcommand{\P}{{\cal{P}}}
\renewcommand{\O}{{\cal{O}}}
\newtheorem{theorem}{Theorem}
\newtheorem{definition}{Definition}
\newtheorem{lemma}{Lemma}
\newtheorem{fact}{Fact}
\newtheorem{observation}{Observation}
\newtheorem{corollary}{Corollary}
\newtheorem{claim}{Claim}

\tikzset{
	every node/.style={
		draw,fill=black,circle,ultra thick,inner sep=.4mm
	}
}
\tikzstyle{vert}=[draw,fill=black,circle,ultra thick,outer sep=.1cm,inner sep=.4mm]
\tikzstyle{sol}=[solid,ultra thick]
\tikzstyle{das}=[dash pattern=on 5pt off 3pt,very thick]
\tikzstyle{snake}=[solid,very thick,decorate,decoration=snake]
\tikzstyle{lbl}=[draw=none, fill=none, inner sep=0mm, rectangle, outer sep=.3cm]

\definecolor{red}{RGB}{146,0,0}
\definecolor{blue}{RGB}{0,109,219}

\begin{document}


\title{Optimal General Matchings\thanks{Partly supported by Polish National Science Center grant UMO-2013/11/B/ST6/01748.}}

\author{Szymon Dudycz\thanks{szymon.dudycz@cs.uni.wroc.pl}, \enskip and
		Katarzyna Paluch\thanks{abraka@cs.uni.wroc.pl}}
		

\maketitle

\begin{abstract}
Given a graph $G=(V,E)$ and for each vertex $v \in V$ a subset $B(v)$ of the set $\{0,1,\ldots, d_G(v)\}$ a $B$-matching of $G$ is any set $F \subseteq E$  such that $d_F(v) \in B(v)$ for each vertex $v$. The general matching problem asks the existence of a $B$-matching in a given graph. A set $B(v)$ is said to have a {\em gap of length} $p$ if there exists a number $k \in B(v)$ such that $k+1, \ldots, k+p \notin B(v)$ and $k+p+1 \in B(v)$. Without any restrictions  the general matching problem is NP-complete.  However, if no set $B(v)$ contains a gap of length greater than $1$, then the problem can be solved in polynomial time and Cornuejols \cite{Cor} presented an algorithm for finding a $B$-matching, if it exists. In this paper we consider a version of the general matching problem, in which we are interested in finding a  $B$-matching having a  maximum (or minimum) number of edges. 

We present the first polynomial time algorithm for  the maximum weight  $B$-matching for the case when no  set $B(v)$ contains a gap of length greater than $1$.

\end{abstract}


\section{Introduction}
Given a graph $G=(V,E)$ and for each vertex $v \in V$ a subset $B(v)$ of the set \\ $\{0,1,\ldots, d_G(v)\}$, where $d_G(v)$ denotes the degree of vertex $v$ in the graph $G$, a $B$-matching of $G$ is any set $F \subseteq E$  such that $d_F(v) \in B(v)$ for each vertex $v$, where $d_F(v)$ denotes the number of edges of $F$ incident to $v$. The general matching problem asks the existence of a $B$-matching in a given graph. Without any restrictions the general matching problem is NP-complete \cite{Lovasz}. 

A set $B(v)$ is said to have a {\em gap of length} $p$ if there exists a natural number $k \in B(v)$ such that $k+1, \ldots, k+p \notin B(v)$ and $k+p+1 \in B(v)$. For the case when no set $B(v)$ contains a gap of length greater than $1$, Lovasz \cite{Lovasz} developed a structural description  and Cornuejols \cite{Cor} presented a polynomial time algorithm for finding a $B$-matching, if it exists. In the maximum/minimum cardinality variant the goal is to find a $B$-matching having a maximum/minimum number of edges. In the weighted version of the problem a weight function $w: E \rightarrow \mathbb N$ is given and the aim is to find a $B$-matching that maximizes or minimizes the sum of the weights of the edges.

Matchings, $b$-matchings and factors are basic combinatorial notions that lie at the foundation of combinatorial optimization. The general matching problem restricted to gaps of at most $1$ is one of the strongest generalizations of matching, that was not proven $\mathcal{NP}$-hard. As such it is of theoretical importance to find a polynomial time algorithm for a maximum/minimum cardinality/weight $B$-matching with gaps at most $1$ or in the case of a maximum weight $B$-matching, to decide if it is $\mathcal{NP}$-hard.
 
{\bf Previous  work} If $B(v)=\{0,1\}$ for each vertex $v$, then a $B$-matching is in fact a {\em matching}, i.e., a set of vertex-disjoint edges.
A {\em perfect matching} is a $B$-matching such that $B(v)=1$ for each vertex $v$. Given a function $b: V \rightarrow \mathbb N$, a $b$-matching is any set 
$F \subseteq E$  such that $d_F(v) \leq b(v)$ for each vertex $v$ and a perfect $b$-matching or a $b$-factor is any set 
$F \subseteq E$  such that $d_F(v) = b(v)$ for each vertex $v$. If in addition to a function $b$ we are also given a function $a: V \rightarrow \mathbb N$, then an $(a,b)$-matching  is  any set 
$F \subseteq E$  such that $a(v) \leq d_F(v) \leq b(v)$ for each vertex $v$.

All these special cases of the general matching problem are well-solved, both in unweighted and weighted versions. For instance, for the maximum weight $b$-matching there exist algorithms with the following running times: $\O(n^2B)$ by Pulleyblank \cite{pulleyblank1973faces}, $\O(n^2m\log B)$ by Marsh \cite{Marsh:1979:MA:909140}, $\O(m^2\log n\log B)$ by Gabow \cite{Gabow1983}, $\O(n^2m+n\log B(m+n\log n))$ and $\O(n^2\log n(m+n\log n))$ by Anstee \cite{Anstee87}, and $\mathcal{\tilde{O}}(W\phi^\omega)$ by Gabow and Sankowski \cite{GabowS13}, where $n=|V|$, $m=|E|$, $B=\max b(v)$, $\phi=\sum b(v)$ and $n^\omega$ is the time required to multiply two $n\times n$ matrices. For a good survey on these problems see \cite{Sch}. 

In the {\em antifactor} problem for each vertex $v$  we have $|\{0,1,\ldots, d_G(v)\} \setminus B(v)| = 1$, meaning that for each vertex there is exactly one degree excluded from the set $B(v)$. Graphs that have an antifactor have been characterized by Lovasz  in \cite{Lovasz1973}.

For the more general case  when no set $B(v)$ contains a gap of length greater than $1$, Cornuejols \cite{Cor} in 1988 presented two solutions to the problem of finding such $B$-matching, if it exists. One uses a reduction to the edge-and-triangle partitioning problem, in which we are given a graph $G=(V,E)$ and a set $T$ of triangles (cycles of length $3$) of $G$ and are to decide if the set of vertices $V$ can be partitioned into sets of cardinality of $2$ and $3$ so that each set of cardinality $2$ is an edge of $E$ and each set of cardinality $3$ is a triangle of $T$. The other is based on an augmenting path approach applied in the modified graph $G'=(V \cup V', E')$ in which  each edge $e$ of $G$ is split with two new vertices  into three edges. For each new vertex $v'$ the set $B(v')$ is defined to be $\{1\}$ and we start from the set $F\subseteq E'$ such that all requirements regarding vertices of $G$ are satisfied, i.e., $d_F(v) \in B(v)$ for each vertex $v \in V$ and for each vertex $v' \in V'$  it is $d_F(v') \leq 1$. Next we aim to gradually augment $F$ so that it also satisfies the requirements regarding new vertices $V'$ and
$d_F(v')=1$ for each $v' \in V'$. In either case, the computed $B$-matching is not guaranteed to be of maximum or minimum cardinality.  A good characterization of graphs that have a $B$-matching \cite{Sebo1993} was provided in 1993 by Seb\H{o}  \cite{Sebo1993}. 

General matchings in bipartite graphs were also studied in terms of their parameterized complexity. 
Gutin et al. showed that for graphs $G=(U\cupdot V, E)$, such that $|B(u)|=1$ for every $u\in U$, there exists a fixed-parameter tractable algorithm parametrized by the size of $V$ \cite{Gutin2012}.

For the optimization variant of the general matching with no gap greater than $1$ Carr and Parekh provided a linear relaxation which is $\frac 12$-integral \cite{Carr2006}.

A $B$-matching is said to be {\em uniform} if each $B(v)$ is either an interval, i.e., has the form $\{a(v), a(v)+1, \ldots, b(v)\}$ for some nonnegative integers $a(v) \leq b(v)$ or an interval intersected with either  even  or odd numbers, i.e., has the form $\{a(v), a(v)+2, \ldots, b(v)\}$  for two nonnegative integers $a(v) \leq b(v)$ such that $b(v)-a(v)$ is even. A maximum/minimum weight uniform $B$-matching problem was shown to be solvable in polynomial time by Szab{\'o} \cite{Szabo}. In the solution to the weighted uniform $B$-matching Szab{\'o} uses the following result of Pap \cite{Pap}. Let $\cal{F}$ be an arbitrary set of odd length cycles of graph $G$, where a single vertex
is considered a cycle of length $1$. A {\em perfect} $\cal{F}$-matching is any set of  cycles and edges of $G$ such that each vertex belongs to exactly one edge or cycle from $\cal{F}$. Pap gave a polynomial time algorithm which minimizes a linear function over the convex hull of perfect $\cal{F}$-matchings.

{\bf Our results} We give the first polynomial time algorithm for  the maximum weight $B$-matching for the case when no  set contains a gap of length greater than $1$.

We provide a structural result for both cardinality and weighted variants, which states that given two $B$-matchings $M$ and $N$, their symmetric difference $M \oplus N= (M \setminus N) \cup (N \setminus M)$  can be decomposed into a set of {\em canonical paths}, a notion which we define precisely later and which plays an analogous role as that of an {\em alternating path} in the context of standard matchings. A path $P$ is alternating with respect to a matching $M$ if its edges alternate between edges of $M$ and edges not belonging to $M$.
Roughly speaking, a canonical path (with respect to a given $B$-matching $M$) consists of a meta-path, that is a sequence of alternating paths, and possibly some number of  meta-cycles attached to the endpoints of this  meta-path. A meta-cycle is a sequence of alternating paths such that the beginning of the first alternating path coincides with the end of the last alternating path in the sequence. After the application of a canonical path ${\cal P}$ to a $B$-matching $M$ we obtain another $B$-matching $M'=M\oplus\P$ such that only the parities of the degrees in $M$ and $M'$ of  the endpoints of ${\cal P}$ are different.

Equipped with this structural result we show how finding a maximum  $B$-matching can be reduced to a series of computations of a maximum/minimum weight
uniform $B$-matching. In fact we prove that in order to verify if a given $B$-matching $M$ has maximum/minimum weight it suffices to check if there exists a uniform $B$-matching of so called {\em neighbouring type} to $M$, whose weight is greater/smaller than that of $M$.

To find a maximum weight $B$-matching, we use the algorithm for maximum cardinality $B$-matching, and run it repeatedly on the same graph, but with varying weight function.

Additionally, we show a very simple reduction of a weighted uniform $B$-matching to a weighted $(a,b)$-matching, which yields a more efficient and simpler algorithm than the one by Szab{\'o}.

{\bf Related work} In the deficiency problems the task consists in finding a matching that is as close as possible to given sets $B(v)$. Hell and Kirkpatrick \cite{HellKirkpatrick} gave an algorithm for finding a minimum deficiency $(a,b)$-matching among all $(0,b)$-matchings, where the deficiency is measured as the sum of differences $a(v)-d(v)$ over all vertices whose degree is not between $a(v)$ and $b(v)$. They also proved that for another measure of deficiency, namely number of vertices whose degree is outside $(a(v),b(v))$, the problem is NP-hard.

Another related problem consists in decomposing a graph into $(a,b)$-matchings - a graph that can be decomposed into $(a,b)$-matchings is called $(a,b)$-factorable. In \cite{Kano} Kano gave a sufficient condition for a graph to be $(2a,2b)$-factorable. Cai \cite{Cai1991} generalized this result to $(2a-1,2b)$, $(2a,2b+1)$ and $(2a-1,2b+1)$ -factorable graphs. Hilton and Wojciechowski showed another sufficient condition for an $(r,r+1)$-factorization of graphs \cite{hilton2005}.

$(a,b)$-matchings were also studied in the stable framework - Biro et al. proved that checking whether a stable $(a,b)$-matching exists is NP-hard \cite{Biro}.

{\bf Organization} In Section \ref{sec:uniform_matching} we present a simple reduction for a uniform $B$-matching. In Section \ref{sec:structure} we introduce the notion of a canonical paths, followed by the proof of the main theorem of our paper. The proof of a key technical lemma is omitted and available in full version\cite{DudyczP17}. In Section \ref{sec:algorithm} we present an algorithm for a maximum $B$-matching and in Section \ref{sec:weighted_alg} for a maximum weight $B$-matching.

\section{Uniform $B$-matching}
\label{sec:uniform_matching}

In this section we show a reduction of a uniform $B$-matching to an $(a,b)$-matching. 

Suppose an instance of a uniform $B$-matching involves a graph $G=(V,E)$ and  for each vertex $v \in V$ a subset $B(v)$ of the set $\{0,1,\ldots, d_G(v)\}$.
We construct a graph $G'=(V , E \cup E')$ and functions $a,b: V \rightarrow N$ as follows. 

If for a vertex $v$ the set $B(v)$  is an interval $\{c(v), c(v)+1, \ldots, d(v)\}$ for some nonnegative integers $c(v) \leq d(v)$, then we set $a(v)=c(v)$ and $b(v)=d(v)$. If for a vertex $v$ the set $B(v)$ has the form $\{c(v), c(v)+2, \ldots, d(v)\}$, i.e., $c(v)$ and $d(v)$ have the same parity and $B(v)$ contains all numbers between $c(v)$ and $d(v)$ of the same parity as $c(v)$, then we add $\frac{d(v)-c(v)}{2}$ loops incident to $v$ and set $a(v)=b(v)=d(v)$. Each loop has weight $0$. Apart from this each edge $e \in E$ has the same weight in $G$ and $G'$. Thus $E'$ consists of some number of loops that are added to each vertex $v$
such that $B(v)$ is not an interval.

\begin{theorem}
\label{thm:uniform_matching}
There is a one-to-one correspondence between $B$-matchings of $G$ and $(a,b)$-matchings of $G'$.
A maximum weight $(a,b)$-matching of $G'$ yields a maximum weight $B$-matching of $G$.
\end{theorem}

After this reduction the number of edges may increase by at most $n^2$ and the number of vertices remains the same. To solve the $(a,b)$-matching we can use an algorithm by Gabow~\cite{Gabow1983}. Its running time on a (multi-)graph with $n$ vertices and $m$ edges is $\sum_{v \in V} b(v) \min(m\log n, n^2)$, which we bound by $n^4$. As the number of vertices does not change in the reduction, a uniform $B$-matching can also be found in time $\mathcal{O}(n^4)$.

\section{Structure of general $B$-matchings}
\label{sec:structure}

In this section we will consider the weighted version of the problem - for the maximum cardinality variant it is enough to set all weights to 1.

Let us first recall and generalise some notions and facts from matching theory. In the case of matchings, it is often convenient to consider the symmetric difference of two matchings.
Given two matchings  $M$ and $N$ the symmetric difference of $M$ and $N$, denoted as $M\oplus N$, is equal to $(M \setminus N) \cup (N \setminus M)$.
The symmetric difference $M\oplus N$ of two matchings $M$ and $N$ can be decomposed into a set of edge-disjoint alternating paths and alternating cycles, 
where a path or cycle is said to be {\em alternating} if its edges belong alternately to $M$ and $N$.  We extend the definiton of an alternating path and cycle to the context of $B$-matchings.

\begin{definition}
Let $M$ be any $B$-matching of $G$. An {\em alternating cycle} (with respect to $M$) is a sequence of edges $P=\\((v_1, v_2), (v_2, v_3), \ldots, (v_{2k-1}, v_{2k}), (v_{2k}, v_1))$ such that

\begin{itemize}
\item for every $i$ such that  $1 \leq i \leq k$  the edge $(v_{2i-1}, v_{2i})$ belongs to $M$,
\item  $(v_{2k},v_1) \notin M$ and for every $i$ such that  $1 \leq i \leq k-1, \ (v_{2i}, v_{2i+1}) \notin M$,
\item each edge of $G$ occurs in $P$ at most once,
\end{itemize}

An {\em alternating path} (with respect to $M$) is a sequence of edges $P=\\((v_1, v_2), (v_2, v_3), \ldots,  (v_{k}, v_{k+1}))$ such that 
\begin{itemize}
\item for every $i$ such that  $1 \leq i \leq k-1$  exactly one of the edges  $(v_{i}, v_{i+1}),(v_{i+1}, v_{i+2})$  belongs to $M$,
\item each edge of $G$ occurs in $P$ at most once,
\item if $v_1=v_{k+1}$, then either both edges $(v_1, v_2)$ and $(v_k, v_1)$ are in $M$, or both are not in $M$.

\end{itemize}

Vertices $v_1$ and $v_{k+1}$ are called the {\em endpoints} of $P$ and edges $(v_1, v_2), (v_{k}, v_{k+1})$ the {\em ending edges} of $P$.

\end{definition}

Notice that the vertices in the definition are not necessarily distinct.

Examples of alternating paths and cycles are shown in Figure \ref{fig:alternating_paths}. Throughout the paper we will draw matched edges using wavy lines, and unmatched edges using straight lines.

\begin{figure}[t]
\centering
\begin{subfigure}{0.35\textwidth}
\centering
\scalebox{0.7}{
\begin{tikzpicture}
	\node (v1) {};
	\node[above right of= v1] (v2) {};
	\node[below right of= v2] (v3) {};
	\node[below left of= v3] (v4) {};

	\draw[snake] (v1) -- (v2);
	\draw[sol] (v2) -- (v3);
	\draw[snake] (v3) -- (v4);
	\draw[sol] (v4) -- (v1);

	\node[draw=none, fill=none, right = 0.5cm of v3] (aux) {};

	\draw (aux) ++(30:1) node (v5) {};
	\draw (aux) ++(-30:1) node (v6) {};
	\draw (v5) ++(-30:1) node (v7) {};
	\draw (v7) ++(30:1) node (v8) {};
	\draw (v7) ++(-30:1) node (v9) {};

	\draw[snake] (v5) -- (v6);
	\draw[sol] (v5) -- (v7);
	\draw[sol] (v6) -- (v7);
	\draw[snake] (v7) -- (v8);
	\draw[snake] (v7) -- (v9);
	\draw[sol] (v8) -- (v9);


\end{tikzpicture}
}
\subcaption{Examples of alternating cycles}
\end{subfigure}
\quad
\begin{subfigure}{0.6\textwidth}
\centering
\scalebox{0.7}{
\begin{tikzpicture}

	\node (v1) {};
	\node[above right of= v1] (v2) {};
	\node[below right of= v2] (v3) {};
	\node[above right of= v3] (v4) {};

	\draw[sol] (v1) -- (v2);
	\draw[snake] (v2) -- (v3);
	\draw[sol] (v3) -- (v4);

	\node[draw=none, fill=none, right = 0.1cm of v4] (aux) {};

	\draw (aux) ++(30:1) node (v5) {};s
	\draw (aux) ++(-30:1) node (v6) {};
	\draw (v5) ++(-30:1) node (v7) {};

	\draw[sol] (v5) -- (v6);
	\draw[snake] (v5) -- (v7);
	\draw[snake] (v6) -- (v7);


	\node[right of= v7] (v8) {};
	\node[right of= v8] (v9) {};
	\node[right of= v9] (v10) {};
	\draw (v10) ++(30:1) node (v11) {};
	\draw (v10) ++(-30:1) node (v12) {};

	\draw[sol] (v8) -- (v9);
	\draw[snake] (v9) -- (v10);
	\draw[sol] (v10) -- (v11);
	\draw[sol] (v10) -- (v12);
	\draw[snake] (v11) -- (v12);

\end{tikzpicture}
}
\subcaption{Examples of alternating paths}
\end{subfigure}
	\caption{}
	\label{fig:alternating_paths}
\end{figure}

The decomposition of the symmetric difference of two $B$-matchings into alternating paths and cycles is not unique. Nevertheless we are interested in {\em maximal} decompositions, i.e., such ones that the concatenation of any two alternating paths from the decomposition
does not result in a new alternating path or cycle.

By {\em applying} an alternating path or cycle $P$ to a $B$-matching $M$ we mean the operation, whose result is $M \oplus P$. We can notice that given any alternating cycle $P$ with respect to a $B$-matching $M$, the set $M'=M \oplus P$ is also a $B$-matching, because $d_{M'}(v)=d_M(v)$ for each vertex $v$. However, it is not true that for every alternating path $P$ with respect to a $B$-matching $M$, $M' =M \oplus P$ is also a $B$-matching.   If $v_1, v_2$ are the endpoints of $P$, then $d_{M'}(v_1)\neq d_M(v_1)$ and $d_{M'}(v_2)\neq d_M(v_2)$, so it may happen that $d_{M'}(v_1) \notin B(v_1)$ or  $d_{M'}(v_2) \notin B(v_2)$.

We observe the following.

\begin{fact}
Given two $B$-matchings $M$ and $N$. Let $D_-$ and $D_+$ denote the sets, respectively, $\{v \in V: d_N(v) < d_M(v)\}$ and $\{v \in V: d_N(v) > d_M(v)\}$ and let $D$ denote $D_- \cup D_+$.
Then any maximal decomposition of $M \oplus N$ has the property that each endpoint of an alternating path from the decomposition belongs to $D$. Also, if $v\in D_-$ is an endpoint of an alternating path $P$, then its ending edge incident to $v$ belongs to $M$ and similarly, every ending edge  of an alternating path $P$ incident to a vertex $v$ in $D_+$ such that $v$ is an endpoint of $P$, belongs to $N$.
\end{fact}

Since the application of an alternating path to a $B$-matching does not necessarily  lead to a new $B$-matching, we need to introduce some generalisation of an alternating path that can be applied in the context of $B$-matchings in a similar way as an alternating path in the context of (standard) matchings.

From alternating paths of a maximal decomposition of the symmetric difference of two $B$-matchings $M$ and $N$ we build {\em meta-paths} and {\em meta-cycles}. Let $P(u,v)$  denote an alternating path with the endpoints $u$ and $v$ (note that $u,v \in D$). A meta-cycle $\cal{C}$ (w.r.t. $M$) is a sequence of alternating paths of the form \\
$(P(v_1, v_2), P(v_2, v_3), \ldots, P(v_k, v_1))$ such that  vertices $v_1, \ldots, v_{k}$ are pairwise distinct. Analogously, a meta-path ${ \cal P}(v_1,v_{k+1})$ (w.r.t. $M$) is a sequence of alternating paths of the form
$(P(v_1, v_2), P(v_2, v_3), \ldots, P(v_k, v_{k+1}))$ such that vertices $v_1, \ldots, v_{k+1}$ are pairwise distinct. Let us note that a meta-cycle may consist of one alternating path of the form $P(v,v)$.

For a vertex $v$ and $k\in B(v)$ let $u_k(v)$ be a  maximum element of $B(v)$, such that $B(v)\cap [k,u_k(v)]$ does not contain an element of different parity than $k$. Because $B(v)$ has a gap of length at most $1$ we obtain that  $B(v)\cap [k,u_k(v)] = \{k, k+2, k+4, \dots, u_k(v)\}$. Also, either $u_k(v) + 1 \in B(v)$ or $u_k(v)$ is a maximum element of $B(v)$, as otherwise we could increase $u_k(v)$. Similarly let us define $l_k(v)$ to be a minimum element of $B(v)$, such that $B(v)\cap [l_k(v),k]$ does not contain an element of different parity than $k$.

We define $B_k(v)$ to be \[B_k(v):=B(v)\cap [l_k(v), u_k(v)] = \{l_k(v), l_k(v)+2, \dots, k, \dots, u_k(v)\}\]
Note that $\{B_k(v)\}_{k\in B(v)}$ is a partition of the set $B(v)$. For a $B$-matching $M$ we also define $B_M(v) = B_{d_{M}(v)}(v)$.

Given a $B$-matching $M$ we say that a $B$-matching $N$ is {\bf \em of the same uniform type} as $M$ if for every vertex $v$ it holds that $d_N(v) \in B_M(v)$.

A $B$-matching $N$ is said to be {\bf \em of neighbouring type} to a $B$-matching $M$  if there exists a set $W$ consisting of at most two vertices such that $\forall w\in W: d_N(w)\notin B_M(w)$ and $\forall v\notin W: d_N(v) \in B_M(v)$ and:
\begin{itemize}
	\item $|W| = 0$, or
	\item $|W| = 2$ and for $w\in W$ $B_M(w)$ and $B_N(w)$ are adjacent, that is $\max(B_M(w)) + 1 = \min(B_N(w))$ or $\max(B_N(w)) + 1 = \min(B_M(w))$, or
	\item $|W| = 1$ and for $w\in W$ there exists $k$, such that $B_k(w)$ is adjacent to both $B_M(w)$ and $B_N(w)$.
\end{itemize}

In other words we allow two vertices to have degree outside of $B_M(v)$, but we place limits on how much they can deviate from that set.

We are now ready to give a definition of a {\bf \em canonical path} - a notion that is going to prove crucial in further analysis and which plays an analogous role as an alternating path in the context of matchings.

\begin{definition}
A canonical path ${\cal S}(v_1, v_k)$ (with respect to a $B$-matching $M$) in a graph $G$ consists of some number of meta-cycles ${\C}_1, {\C}_2, \ldots, {\C}_p$ incident to a vertex $v_1$, some number of meta-cycles ${\C'}_1, {\C'}_2, \ldots, {\C'}_q$ incident to $v_k$ 
and in case $v_1 \neq v_k$ - of a meta-path $\P(v_1,v_{k})$ such that the application of all meta-cycles ${\C}_1, {\C}_2, \ldots, {\C}_p$, ${\C'}_1, {\C'}_2, \ldots, {\C'}_q$ and the
meta-path $\P(v_1, v_k)$ to $M$ results in a $B$-matching of neighbouring type to $M$.

\end{definition}

Two variants of canonical path - with one endpoint or two endpoints - correspond to different cases in the definition of neighbouring type. Namely, if $v_1 \neq v_k$, then set $W=\{v_1,v_k\}$. Otherwise $v_1=v_k$ and $W=\{v_1\}$ or $W=\emptyset$. The examples of these cases are presented on Figure \ref{fig:can_path_types}.

\begin{figure}[t]
	\centering
\centering
\begin{subfigure}{0.3\textwidth}
\centering
\scalebox{0.8}{
	\begin{tikzpicture}
	\node[red] (v1) {};
	\node[red, above right of= v1] (v2) {};
	\node[red, below right of= v2] (v3) {};
	\node[red, below left of= v3] (v4) {};

	\draw[snake] (v1) -- (v2);
	\draw[sol] (v2) -- (v3);
	\draw[snake] (v3) -- (v4);
	\draw[sol] (v4) -- (v1);

	\end{tikzpicture}
	}
	\subcaption{$W=\emptyset$ and $M\oplus N$ is an alternating cycle.}
\end{subfigure}
\quad
\begin{subfigure}{0.65\textwidth}
\centering
\scalebox{0.8}{
	\begin{tikzpicture}

		\node[label=below right:$v$] (v7) {};

		\draw (v7) ++(120:1.2cm) node[red] (v8) {};
		\draw (v7) ++(60:1.2cm) node[red] (v9) {};
		\draw[sol] (v7) -- (v8);
		\draw[sol] (v7) -- (v9);
		\draw[snake] (v8) -- (v9);

		\draw[] (v7) ++ (20:1.2cm) node[red] (v10) {};
		\draw (v10) ++ (-30:1cm) node[red] (v11) {};
		\draw (v11) ++ (-90:1cm) node[blue] (v12) {};
		\draw (v12) ++ (-150:1cm) node[red] (v13) {};
		\draw (v13) ++ (-210:1cm) node[red] (v14) {};

		\draw[sol] (v7) -- (v10);
		\draw[snake] (v10) -- (v11);
		\draw[sol] (v11) -- (v12);
		\draw[sol] (v12) -- (v13);
		\draw[snake] (v13) -- (v14);
		\draw[sol] (v14) -- (v7);

		\draw (v7) ++ (240:1cm) node[blue] (v15) {};
		\draw (v15) ++ (215:1cm) node[red] (v16) {};
		\draw (v16) ++ (195:1cm) node[red] (v17) {};
		\draw (v17) ++ (165:1cm) node[blue] (v18) {};
		\draw (v18) ++ (100:1cm) node[red] (v19) {};
		\draw (v19) ++ (55:1cm) node[red] (v20) {};
		\draw (v20) ++ (15:1cm) node[red] (v21) {};
		\draw (v21) ++ (-10:1cm) node[red] (v22) {};

		\draw[sol] (v7) -- (v15);
		\draw[sol] (v15) -- (v16);
		\draw[snake] (v16) -- (v17);
		\draw[sol] (v17) -- (v18);
		\draw[sol] (v18) -- (v19);
		\draw[snake] (v19) -- (v20);
		\draw[sol] (v20) -- (v21);
		\draw[snake] (v21) -- (v22);
		\draw[sol] (v22) -- (v7);

	\end{tikzpicture}
	}
	\subcaption{$B_v=\{0,1,3,5,6\}$. Then $W=\{v\}$ and $M\oplus N$ is a canonical path with one endpoint.}

\end{subfigure}

\vspace{2em}

\begin{subfigure}{0.3\textwidth}
\centering
\scalebox{0.8}{
	\begin{tikzpicture}
	\node[label=below:$v$] (v1) {};
	\node[above right of= v1, red] (v2) {};
	\node[above of= v2, red] (v3) {};
	\node[above left of= v1, red] (v4) {};
	\node[above of= v4, red] (v5) {};

	\draw[sol] (v1) -- (v2);
	\draw[snake] (v2) -- (v3);
	\draw[sol] (v3) -- (v5);
	\draw[snake] (v5) -- (v4);
	\draw[sol] (v1) -- (v4);
	\end{tikzpicture}
	}
	\subcaption{If $B_v=\{0,2\}$ then $W=\emptyset$. If $B_v=\{0,1,2\}$ then $W=\{v\}$. In both cases it is a canonical path with one endpoint.}
\end{subfigure}
\quad
\begin{subfigure}{0.65\textwidth}
\centering
\scalebox{0.8}{
\begin{tikzpicture}
	\node[label=below:$u$] (v0) {};
	\node[right of= v0, red] (v1) {};
	\node[right of= v1, red] (v2) {};
	\node[right of= v2, red] (v3) {};
	\node[right of= v3, red] (v4) {};
	\node[right of= v4, blue] (v5) {};
	\node[right of= v5, red] (v6) {};
	\node[right of= v6, label=below left:$v$] (v7) {};

	\draw (v7) ++(120:1cm) node[red] (v8) {};
	\draw (v7) ++(60:1cm) node[red] (v9) {};
	\draw[sol] (v7) -- (v8);
	\draw[sol] (v7) -- (v9);
	\draw[snake] (v8) -- (v9);

	\draw[] (v7) ++ (20:1.2cm) node[red] (v10) {};
	\draw (v10) ++ (-30:1cm) node[red] (v11) {};
	\draw (v11) ++ (-90:1cm) node[blue] (v12) {};
	\draw (v12) ++ (-150:1cm) node[red] (v13) {};
	\draw (v13) ++ (-210:1cm) node[red] (v14) {};

	\draw[sol] (v7) -- (v10);
	\draw[snake] (v10) -- (v11);
	\draw[sol] (v11) -- (v12);
	\draw[sol] (v12) -- (v13);
	\draw[snake] (v13) -- (v14);
	\draw[sol] (v14) -- (v7);

	\draw[snake] (v0) -- (v1);
	\draw[sol] (v1) -- (v2);
	\draw[snake] (v2) -- (v3);
	\draw[sol] (v3) -- (v4);
	\draw[snake] (v4) -- (v5);
	\draw[snake] (v5) -- (v6);
	\draw[sol] (v6) -- (v7);



\end{tikzpicture}
}
\subcaption{$B_u=\{0,1\}$ and $B_v=\{0,1,3,5\}$. Then $W=\{u,v\}$ and $M\oplus N$ is a canonical path with two endpoints.}
\end{subfigure}

	\caption{Examples of matchings of neighbouring types. Solid edges belong to matching $M$ and wavy edges belong to matching $N$. For every red vertex $w$ $B_w=\{1\}$ and for every blue vertex $z$ $B_z=\{0,2\}$.}
	\label{fig:can_path_types}

\end{figure}

We will often refer to the weight of  a canonical path - that is the effect its application has on a $B$-matching $M$. More precisely, for a canonical path $\S$ $w_M(\S) = w(M\oplus \S) - w(M) = \sum_{e\in \S\setminus M} w(e) - \sum_{e\in \S\cap M} w(e)$. Observe that for two edge-disjoint canonical paths $\S_1$ and $\S_2$ we have that $w_M(\S_1) = w_{M\oplus \S_2}(\S_1)$. We will usually write $w(\S)$ instead of $w_M(\S)$ when the choice of $M$ is clear. Also, when constructing new canonical paths, we will use the notion of a fine vertex - we say that a vertex $v$  is {\em fine in $\S$} if the number of edges incident to $v$ in $M \oplus \S$ belongs to $B(v)$ and {\em wrong} otherwise. We say that an endpoint of $\S$ is fine (wrong) if it is fine (wrong) in $\S$. We will say that a path (or cycle) is {\em positive} if its weight is positive.

In our algorithm we want to subsequently find and apply positive weight canonical paths until a $B$-matching is optimal.
Let us start by showing that it is necessary to consider canonical paths, that is that it may happen that a $B$-matching is not optimal, but there is no meta-path or meta-cycle augmenting it (i.e. increasing its size). Consider an unweighted graph in Figure \ref{fig:ex_meta} and let $B(v) = \{0,1,3,5\}$, $B(u) = \{0,1\}$, $B(w) = \{0,2\}$ and $B(t) = \{0,2\}$. For every other vertex $x$ let $B(x) = \{1\}$. Then we cannot apply any of the meta-cycles incident to $v$, because the degree of $v$ would be $2$. On the other hand applying the meta-path decreases the size of the $B$-matching. Hence we need to apply both meta-cycles and the meta-path at the same time (which together form a canonical path) to obtain a feasible $B$-matching of greater size.

\begin{figure}[t]
	\centering

\begin{tikzpicture}
	\node[label=below:$u$] (v0) {};
	\node[right of= v0] (v1) {};
	\node[right of= v1] (v2) {};
	\node[right of= v2] (v3) {};
	\node[right of= v3] (v4) {};
	\node[right of= v4, label=below:$w$] (v5) {};
	\node[right of= v5] (v6) {};
	\node[right of= v6, label=below left:$v$] (v7) {};

	\draw (v7) ++(120:1cm) node (v8) {};
	\draw (v7) ++(60:1cm) node (v9) {};
	\draw[sol] (v7) -- (v8);
	\draw[sol] (v7) -- (v9);
	\draw[snake] (v8) -- (v9);

	\draw[] (v7) ++ (20:1.2cm) node (v10) {};
	\draw (v10) ++ (-30:1cm) node (v11) {};
	\draw (v11) ++ (-90:1cm) node[label=below:$t$] (v12) {};
	\draw (v12) ++ (-150:1cm) node (v13) {};
	\draw (v13) ++ (-210:1cm) node (v14) {};

	\draw[sol] (v7) -- (v10);
	\draw[snake] (v10) -- (v11);
	\draw[sol] (v11) -- (v12);
	\draw[sol] (v12) -- (v13);
	\draw[snake] (v13) -- (v14);
	\draw[sol] (v14) -- (v7);

	\draw[snake] (v0) -- (v1);
	\draw[sol] (v1) -- (v2);
	\draw[snake] (v2) -- (v3);
	\draw[sol] (v3) -- (v4);
	\draw[snake] (v4) -- (v5);
	\draw[snake] (v5) -- (v6);
	\draw[sol] (v6) -- (v7);



\end{tikzpicture}

	\caption{Example of a $B$-matching, which is not optimal, but there is no meta-path or meta-cycle improving it.}
	\label{fig:ex_meta}

\end{figure}

In the remainder of this section we will prove Theorem \ref{thm:canonical_path}, which states that if a $B$-matching $M$ is not optimal, then there exists a canonical path improving it, i.e., such one that its application to $M$ gives rise to a $B$-matching of greater weight. The outline of the proof is as follows. First, in Lemma \ref{lem:path_sequence} we prove  that any $B$-matching can be transformed into an optimal one by a sequence of canonical paths. As an optimal $B$-matching has greater weight, at least one of those paths has positive weight. Next, in Lemma \ref{podst} we prove that we can change the order of the canonical paths in such a way that  positive weight paths  occur earlier in the sequence. The section finishes with the proof of Theorem \ref{thm:canonical_path}, in which we apply a key technical Lemma \ref{podst} to show that we may assume that already the first path in the sequence  has positive weight.

In the proof we will use a more restricted version of a canonical path. In the example above we have seen that we cannot consider only minimal (with respect to inclusion) canonical paths. Therefore, we introduce another notion, similar to a minimal canonical path but taking into account the weight of a path.

\begin{definition}
We say that ${\S}$ is a basic (canonical) path if it is a canonical path and for no proper subset ${\S}'\subsetneq {\S}$ $\S'$ is a canonical path such that  either $w({\S}') \geq w(\S)$ or $w(\S') > 0$.  
\end{definition}

\begin{observation}
\label{obs:basic_path}
Let $M$ be a $B$-matching. If there exists a canonical path $\S$ w.r.t. $M$, then there exists a basic canonical path $\S'\subseteq \S$ w.r.t $M$.
\end{observation}

\begin{lemma}\label{decomp}
\label{lem:path_sequence}
Let $M,N$ be two $B$-matchings. Then there exists a sequence  ${\cal S}_1, {\cal S}_2, \ldots, {\cal S}_k$ and a set of alternating cycles 
$C_1, C_2, \ldots, C_l$  that satisfy the following.
\begin{enumerate}

\item Let $M_0$ denote $M \oplus \bigcup_{i=1}^l C_i$. For each $i$ such that $0 < i \leq k$  ${\cal S}_i$ is a basic canonical path with respect to $M_{i-1}$ and $M_i = M_{i-1} \oplus {\cal S}_i$.   
 Also, $M_k=N$.

\item $M\oplus N = \bigcup_{i=1}^k {\cal S}_i \cup \bigcup_{i=1}^l C_i$, where every two elements of the set  \\ $\{\S_1, \ldots, \S_k, C_1, \ldots, C_l\}$ are edge-disjoint.
\end{enumerate}

\end{lemma}

\begin{proof}
Let us consider some fixed maximal decomposition of $M \oplus N$.
Let $C_1, C_2, $$\ldots, $$C_l$ denote all alternating cycles of this decomposition. By $M_0$ we  denote $M  \oplus \bigcup_{i=1}^l C_i$.

If $d_M(v)=d_N(v)$ for every vertex $v$, then $M \oplus N$ consists solely of alternating cycles $C_1, C_2, \ldots C_l$ and $M_0=N$ and we are done.

The maximal decomposition of $M_0 \oplus N$ consists only of alternating paths.
The {\em distance} of two $B$-matchings $M$ and $N$, denoted as $dist(M,N)$, is defined as \[dist(M,N)=\sum_{v \in V} |d_N(v)-d_M(v)|\] In the distance of two $B$-matchings
it is enough to consider the vertices belonging to $D$, i.e., $dist(M,N)= \sum_{v \in D} |d_N(v)-d_M(v)|$.

Let $M_0$ and $N$ be two matchings such that the set $D$ corresponding to them is not empty, i.e. there exists a vertex $v$ such that $d_{M_0}(v)\neq d_N(v)$
and hence $dist(M_0,N) >0$. 
We show how to construct some canonical path $\cal{S}$ with respect to $M_0$ such that the $B$-matching $M_1=M_0 \oplus \cal{S}$ satisfies: $D(M_1,N) \subseteq D(M_0,N)$,  $D_{-}(M_1,N) \subseteq D_{-}(M_0,N), D_{+}(M_1,N) \subseteq D_{+}(M_0,N)$ and  $dist(M_1,N) < dist(M_0,N)$.

 We start from any alternating path $P$ that belongs to a maximal decomposition of $M_0 \oplus N$. $P$ may have two different endpoints or one endpoint. If $P$ is not a canonical path, then it means that after its application for at least one of its endpoints $v_1$ or $v_2$ it holds that $d_{M_0 \oplus P}(v_i) \notin B(v_i)$, where $i\in \{1,2\}$. We can notice that apart from this $P$ satisfies all the other
conditions of a canonical path. We are going to gradually extend $P$ so that we obtain $\cal{S}$ that is a canonical path. At each stage of the construction
the candidate $\cal{S}$ for a canonical path has all the properties of a canonical path except for the fact that for one or two of its endpoints
it holds that $d_{M_0 \oplus \cal{S}}(v_i) \notin B(v_i)$, where $i\in \{1,2\}$. 

Note that in $\S$ both endpoints have degree one. If $v_i$ is not fine in $\S$ it means that either $B(v_i)$  contains $d_{M_0}(v_i)$ and $d_{M_0}(v_i) + 2$, but  does not contain $d_{M_0}(v_i)+1$ ($v_i \in D_{+})$, or $B(v_i)$  contains $d_{M_0}(v_i)$ and $d_{M_0}(v_i) - 2$, but  does not contain $d_{M_0}(v_i)-1$ ($v_i \in D_{-})$. Then if we add another alternating path starting at $v_i$, it will cease to be an endpoint of $\S$ and its degree will belong to $B_{M_0}(v_i)$. This will be true at each step of our construction - a vertex $v$ that  is not an endpoint satisfies $d_{M_0 \oplus \S}(v) \in B_{M_0}(v)$. Another invariant that will be maintained during the construction is the following: if there are two endpoints of $\S$ their degrees will be odd in $\S$, and if the two endpoints join into one (thus $v_1=v_2$), then their degree is even in $\S$. 

Assume then that we have some candidate path with one endpoint $v_1$ or two endpoints $v_1, v_2$, which is not a canonical path, so $d_{M_0 \oplus \cal{S}}(v_1) \notin B(v_1)$. Since $N$ is a $B$-matching there exists an alternating path $P'$ in the maximal decomposition of $ (M_0 \oplus \cal{S}) \oplus N$ with one endpoint $v_1$. This path has the property that either $P$ and $\S$ both diminish the number of edges incident to $v_1$, or they both increase the number of edges incident to $M_0$, or our alternating paths would not be maximal. After adding $P$ to $\cal{S}$ the following things may happen:
\begin{enumerate}
\item $P$ has two different endpoints $v_1, v_3$. Then vertex $v_1$ is  fine in $\S \cup P$. If $v_3$ is not an endpoint of any alternating path belonging to $\cal{S}$, then $v_3$ is a new endpoint of $\cal{S} \cup P$ and either (i) $v_3$ is fine in $\S \cup P$ and  we have decreased the number of wrong endpoints by one or (ii)  $v_3$ is wrong in $M \oplus (\S \cup P)$  and the number of wrong endpoints of $\S \cup P$ is the same as the number of wrong endpoints of $\S$  and we continue the process treating $\S \cup P$ as the new candidate for a canonical path. If $v_3$ is an endpoint of some alternating path belonging  to $\cal{S}$, then we have created
a new meta-cycle $\cal{C}$ incident to $v_3$. If $v_3$ is fine in $\S \cup \C$, then we decreased the number of wrong endpoints. If $v_3$ is fine in $\C$ then $\C$ is a canonical path with respect to $M_0$. Otherwise it means that $d_{M_0}(v_3) + 2 \notin B(v_3)$, so $v_3$ must be the other endpoint of $\S$. In this case we have only one wrong endpoint left, $v_3$, and we continue extending $\S$ from $v_3$. Note that now that two endpoints have joined in $v_3$, we seemingly have only one endpoint. However, after the addition of an alternating path with two endpoints $v_3$ and $v'$, $\S$ will have two endpoints - $v_3$ and $v'$, where $v_3$ is fine.

\item $P$ has one endpoint $v_1$. If $v_1$ is fine in $\cal{S} \cup \cal{C}$, then we have decreased the number of wrong endpoints of a candidate for a canonical path. Otherwise if
$P$ is a canonical path we are done. The only case left is when $v_1$ is not fine but $d_{M_0}(v_1) + 2\notin B(v_1)$. This may only happen if both endpoints of $\S$ are the same vertex and then we continue extending $\S$ with only one wrong endpoint left.
\end{enumerate}

That way we have constructed a canonical path $\S$ w.r.t. $M$. By Observation \ref{obs:basic_path} it means that there exists a basic canonical path $\S'$.
We can continue finding canonical paths in the same way, this time in $(M_0\oplus \S')\oplus N$. Each such basic canonical path decreases the distance between $M$ and $N$, which means  that way we can decompose $M_0 \oplus N$ into a finite number of basic canonical paths. 

\end{proof}

Now we are ready to state the key technical lemma.

\begin{restatable}[]{lemma}{technicallemma}\label{podst}
Let $M$ and $N$ be two $B$-matchings, such that $w(M) < w(N)$. Let $\Q$ be a basic canonical path w.r.t. $M$ contained in $M \oplus N$ and $\R$ a basic canonical path w.r.t. $M\oplus \Q$ such that $w(\Q) \leq 0$ and $w(\R) > 0$. Then there exists a canonical path $\T$ w.r.t. $M$ such that $w(\T) > w(\Q)$.
\end{restatable}

We defer the proof of this lemma to the full version of this paper~\cite{DudyczP17} and  let us focus on its consequences.

\begin{theorem}
\label{thm:canonical_path}
If there exists a $B$-matching of greater weight than $M$, then there exists a $B$-matching of greater weight than $M$ that is of the same uniform type as $M$ or that is of neighbouring type to $M$.
\end{theorem}

\begin{proof}
Suppose that there does not exist a $B$-matching $M'$ of the same uniform type as $M$ and with greater weight than $M$ but there exists a $B$-matching $N$ having greater weight than $M$.

By Lemma \ref{decomp} we know that there exists a sequence of basic canonical paths ${\cal S}_1, {\cal S}_2, \ldots, {\cal S}_k$ and  a set of alternating cycles 
$C_1, C_2, \ldots, C_l$ such that $M\oplus N = \bigcup_{i=1}^k {\cal S}_i \cup \bigcup_{i=1}^l C_i$. The weight of $N$ satisfies $w(N)=w(M)+ \sum_{i=1}^l w(C_i)+
\sum_{i=1}^k w({\cal S}_i)$. Since $w(N)>w(M)$ there exists an alternating cycle $C_i$ or a canonical paths $\S_i$ with positive weight.

We may, however, observe, that if some alternating cycle $C_i$ has positive weight, then $M \oplus C_i$ is of the same uniform type as $M$ and has greater weight than $M$. As alternating cycles do not change the degree of any vertex, we may apply them after canonical paths. Therefore, let $N' = M \oplus \bigcup_{i=1}^k {\S}_i$ and note that it is also a $B$-matching, as $\forall v d_N'(v) = d_N(v)$. Its weight, however, is greater than the weight of $N$, because we omitted  negative weight alternating cycles. Therefore, we can assume that the decomposition of $M\oplus N$ does not contain any alternating cycles. 

By Lemma \ref{decomp} there exists some sequence of basic canonical paths that forms a decomposition of $M \oplus N$, but it is not necessarily unique. From all such sequences let us choose that one, in which ${\cal S}_1$ has maximum weight. Let $M_1$ denote $M \oplus \S_1$. For each $i>1$, ${\cal S}_i$ is a basic canonical path with respect to $M_{i-1}$ of maximum weight and $M_i=M_{i-1} \oplus \S_i$. 

Note that when choosing $\S_i$ of maximum weight, we will always be able to complete the sequence of canonical paths, because $M_i$ is a $B$-matching and thus we can apply Lemma \ref{decomp}.

Some basic canonical path ${\cal S}_i$ must of course have positive weight. Let $i$ be the smallest such index. If $i=1$, then we are
done. Assume then, that $i>1$. 

It means that $\S_i$ has positive weight and $w({\cal S}_{i-1}) \leq 0$. Then, by Lemma \ref{podst} and Observation \ref{obs:basic_path}, there exists a basic canonical path ${\cal S'}_{i-1}$ with respect to $M_{i-2}$ such that $w(\S'_{i-1}) > w(\S_{i-1})$, which contradicts the properties of our decomposition, because instead of adding ${\cal S}_{i-1}$, we would choose ${\cal S'}_{i-1}$.

Such argument cannot be applied only if the weight of ${\cal S}_1$ is already positive, which shows that the claim of the theorem is correct.
\end{proof}

\section{Algorithm for computing a maximum cardinality $B$-matching}
\label{sec:algorithm}

In this section we will show the algorithmic consequences of Theorem \ref{thm:canonical_path}, namely we will present a polynomial time algorithm for a maximum cardinality $B$-matching.

First, let us assume that we have some $B$-matching $M$. We want to be able to either verify that it is maximum or find a $B$-matching of greater cardinality. 
According to Theorem \ref{thm:canonical_path}, $M$ is not maximum if and only if there exists a larger $B$-matching $M'$ such that at most two vertices' degrees are not in $B_M(v)$. Therefore, we can consider all possible sets of at most two vertices, whose degrees would not be restricted to $B_M(v)$. For the rest of vertices we allow them to have any degree in $B_M(v)$. This is an instance of a uniform $B$-matching, hence we use Theorem \ref{thm:uniform_matching} to solve it.

This approach requires solving $O(n^2)$ instances of a maximum weight
uniform $B$-matching problem. 

In order to  find a maximum cardinality $B$-matching we start by running Cornuejols' algorithm, which finds any $B$-matching or verifies that the graph does not have admit a $B$-matching. Then we subsequently augment this matching until it is maximum. The size of a maximum matching can be bounded by the number of edges in the graph, thus the total complexity is $\mathcal{O}(m n^6)$.

This algorithm can be also used for finding a maximum weight $B$-matching, however, since the value maximum weight $B$-matchings can be bounded only by $mW$, where $W = \max w(e)$, the algorithm becomes pseudopolynomial.

\noindent \fbox{
\begin{minipage}[t]{0.956\textwidth}
\vspace{0.5cm}
{\bf \em \hspace{0.5cm} Algorithm 1. Max $B$-Matching}
\vspace{0.5cm}
\begin{enumerate}
\item Let $M$ be any $B$-matching (e.g. from Cornuejols' algorithm)
\item {\bf while}  there exists a $B$-matching $M'$ of neighbouring type to $M$ with cardinality greater than that of $M$ {\bf do:}

\hspace{1cm} $M \leftarrow M'$ 
\item Output $M$
\end{enumerate}
\vspace{0.5cm}
\end{minipage}
}

\section{Algorithm for weighted $B$-matching}
\label{sec:weighted_alg}

In this section we will show how to use algorithm from previous section to solve weighted $B$-matching.

Let us start by showing why algorithm for maximum cardinality is too slow for the weighted version. A canonical path may increase weight of a matching by $1$, and the weight of a maximum $B$-matching can be bounded only by $mW$, which gives a pseudopolynomial time algorithm.

We would like, however, to force the algorithm to start from heavy edges, to make faster progress. To do this, we will be changing the weights and running repeatedly the algorithm for maximum cardinality (the graph and sets $B(v)$ will remain unchanged throughout the algorithm).

In $i$-th iteration our weights will consist of $i$ most significant bits of original weights. Then, we will improve the matching from the previous iteration using algorithm for maximum cardinality.

More formally, let $w:E \rightarrow \mathbb{N}$ be weights of the edges, let $W= \max w(e)$ and let $l$ be its length (so $l=\lceil \log(W) \rceil$). Then let $w_i = \lfloor\frac{w}{2^{l-i}}\rfloor$ be the weights in $i$-th iteration. 

As $M_i$ we will denote the maximum $B$-matching after $i$-th iteration (the starting matching, $M_0$, can be any feasible $B$-matching). The pseudocode of the algorithm is presented below.

Now, we want to show that in the $i$-th iteration $B$-matching will be improved at most $m$ times.

\begin{lemma}
$w_{i+1}(M_{i+1}) \leq w_{i+1}(M_i) + m$
\end{lemma}
\begin{proof}
Assume otherwise that $w_{i+1}(M_{i+1}) > w_{i+1}(M_i) + m$.

 We know that $w_{i+1}(M_i) \geq 2 * w_i(M_i)$ and $w_{i+1}(M_{i+1}) \leq 2 * w_i(M_{i+1}) + m$, where the last inequality follows from the fact, that there are at most $m$ edges in $M_{i+1}$, and by setting the least significant bit of weight to 0, weight of the edge can decrease by at most 1 (then it is halved, when the least significant bit is removed).

So it follows, that $w_i(M_{i+1}) > w_i(M_i)$, which contradicts optimality of $M_i$.
\end{proof}

As the weights are natural numbers, each time the weight of matching is increased by at least 1, which proves that there are at most $m$ iterations.

Therefore, the total running time of the algorithm is $\mathcal{O}(\log(W) m n^6)$.

\noindent \fbox{
\begin{minipage}[t]{0.956\textwidth}
\vspace{0.5cm}
{\bf \em \hspace{0.5cm} Algorithm 2. Max Weight $B$-Matching}
\vspace{0.5cm}
\begin{enumerate}
\item  Let $M$ be any $B$-matching
\item {\bf for} $i=1$ to $\lceil \log(W) \rceil$:

	\hspace{1cm} Run Algorithm 1 on $M$ w.r.t. weights $w_i$

\item Output $M$
\end{enumerate}
\vspace{0.5cm}
\end{minipage}
}

\bibliographystyle{plain}
\bibliography{optfactor}

\begin{thebibliography}{10}

\bibitem{Anstee87}
Richard~P. Anstee.
\newblock A polynomial algorithm for b-matchings: An alternative approach.
\newblock {\em Inf. Process. Lett.}, 24(3):153--157, 1987.

\bibitem{Biro}
Péter Biró, Tamás Fleiner, Robert~W. Irving, and David~F. Manlove.
\newblock The college admissions problem with lower and common quotas.
\newblock {\em Theoretical Computer Science}, 411(34):3136 -- 3153, 2010.

\bibitem{Carr2006}
Robert Carr and Ojas Parekh.
\newblock A 12-integral relaxation for the a-matching problem.
\newblock {\em Operations Research Letters}, 34(4):445 -- 450, 2006.

\bibitem{Cor}
Gérard Cornuéjols.
\newblock General factors of graphs.
\newblock {\em Journal of Combinatorial Theory, Series B}, 45(2):185 -- 198,
  1988.

\bibitem{DudyczP17}
Szymon Dudycz and Katarzyna~E. Paluch.
\newblock Optimal general matchings.
\newblock {\em CoRR}, abs/1706.07418, 2017.

\bibitem{Gabow1983}
Harold~N. Gabow.
\newblock An efficient reduction technique for degree-constrained subgraph and
  bidirected network flow problems.
\newblock In {\em Proceedings of the Fifteenth Annual ACM Symposium on Theory
  of Computing}, STOC '83, pages 448--456, New York, NY, USA, 1983. ACM.

\bibitem{GabowS13}
Harold~N. Gabow and Piotr Sankowski.
\newblock Algebraic algorithms for b-matching, shortest undirected paths, and
  f-factors.
\newblock In {\em 54th Annual {IEEE} Symposium on Foundations of Computer
  Science, {FOCS} 2013, 26-29 October, 2013, Berkeley, CA, {USA}}, pages
  137--146, 2013.

\bibitem{Gutin2012}
Gregory Gutin, Eun~Jung Kim, Arezou Soleimanfallah, Stefan Szeider, and Anders
  Yeo.
\newblock Parameterized complexity results for general factors in bipartite
  graphs with an application to constraint programming.
\newblock {\em Algorithmica}, 64(1):112--125, 2012.

\bibitem{HellKirkpatrick}
Pavol Hell and David~G. Kirkpatrick.
\newblock Algorithms for degree constrained graph factors of minimum
  deficiency.
\newblock {\em J. Algorithms}, 14(1):115--138, 1993.

\bibitem{hilton2005}
A.~J.~W. Hilton and Jerzy Wojciechowski.
\newblock Semiregular factorization of simple graphs.
\newblock {\em AKCE Int. J. Graphs Comb}, 2(1):57--62, 2005.

\bibitem{Kano}
Mikio Kano.
\newblock [a,b]-factorization of a graph.
\newblock {\em Journal of Graph Theory}, 9(1):129--146, 1985.

\bibitem{Lovasz1973}
L{\'{a}}szl{\'{o}} Lov{\'a}sz.
\newblock Antifactors of graphs.
\newblock {\em Periodica Mathematica Hungarica}, 4(2):121--123, 1973.

\bibitem{Lovasz}
L{\'{a}}szl{\'{o}} Lovász.
\newblock The factorization of graphs. ii.
\newblock {\em Acta Mathematica Hungarica}, 23(1-2):223--246, 1972.

\bibitem{Cai1991}
Cai Mao-Cheng.
\newblock [a,b]-factorizations of graphs.
\newblock {\em Journal of Graph Theory}, 15(3):283--301, 1991.

\bibitem{Marsh:1979:MA:909140}
Alfred~Burton Marsh, III.
\newblock {\em Matching Algorithms.}
\newblock PhD thesis, The Johns Hopkins University, 1979.

\bibitem{Pap}
Gyula Pap.
\newblock A {TDI} description of restricted 2-matching polytopes.
\newblock In {\em Integer Programming and Combinatorial Optimization, 10th
  International {IPCO} Conference, New York, NY, USA, June 7-11, 2004,
  Proceedings}, pages 139--151, 2004.

\bibitem{pulleyblank1973faces}
William~R. Pulleyblank.
\newblock {\em Faces of Matching Polyhedra, Univ. of Waterloo, Dept.
  Combinatorics and Optimization}.
\newblock PhD thesis, University of Waterloo, 1973.

\bibitem{Sch}
Alexander Schrijver.
\newblock {\em Combinatorial Optimization: Polyhedra and Efficiency},
  volume~24.
\newblock Springer Science \& Business Media, 2002.

\bibitem{Sebo1993}
Andr{\'{a}}s Seb{\"{o}}.
\newblock General antifactors of graphs.
\newblock {\em J. Comb. Theory, Ser. {B}}, 58(2):174--184, 1993.

\bibitem{Szabo}
Jácint Szabó.
\newblock Good characterizations for some degree constrained subgraphs.
\newblock {\em Journal of Combinatorial Theory, Series B}, 99(2):436 -- 446,
  2009.

\end{thebibliography}

\begin{appendix}

\section{Structure and properties of a basic canonical path}
\label{sec:lemma_proof}

\label{sec:key_lemma}

In this Section we will prove Lemma \ref{podst}. Let us start with some notation we will use throughout this section. Each path and cycle in this section  denotes a meta-path and a meta-cycle. Also we will use the relative notation of degrees. If $M$ is a $B$-matching and $v$ some vertex, then we will use the set $\{d - d_M(v) : \forall d \in B(v)\}$. Particularly, we will use $0$ to denote the current degree.

We say that a vertex $v$ is {\bf odd w.r.t. $M$} if $deg_{M}(v) +1 \in B(v)$ and {\bf even w.r.t. $M$} otherwise. 
We will often omit $M$ and say that a vertex $v$ is odd (even) if it is odd (even) w.r.t. $M$. We will also say that a vertex $v$ is odd (even) w.r.t. $\S$ if it is odd (even) w.r.t. $M\oplus \S$. 

For any canonical path $\S$ w.r.t. $M$ we will also assume that if any vertex $v$ is in $D$, then it is in $D_+$. That is because the case when $v\in D_-$ is completely symmetrical, so we will avoid repeating each argument twice.

Now let us state the following observation which is a consequence of the definition of  a $B$-matching of neighbouring type.

\begin{observation}
\label{obs:degrees_canonical_path}
Let $M$ be a $B$-matching and let $\S$ be a canonical path. Let $N=M\oplus \S$ and $v, w$ be the endpoints of $\S$. Then:
\begin{enumerate}
	\item For each vertex $u$ other than $v$ and $w$ $\{0, 2, \dots, d_{N\oplus M}(u)\} \subseteq B(u)$.
	\item If $v$ and $w$ are distinct, then for $u \in \{v,w\}$ there is some $k\in \{0,2,\dots, d_{M\oplus N}(u) - 1\}$ such that $\{0,2,4,\dots, k, k+1, k+3, \dots, d_{M\oplus N}(u)\} \subseteq B(u)$.
	\item If $v = w$ then $\{0, 2, \dots, d_{M\oplus N}(v)\}\subseteq B(v)$ or there are $k_1 \in \{0, 2, \dots, d_{M\oplus N}(v)-2\}$ and $k_2 \in \{k_1+1, k_1+3,\dots, d_{M\oplus N}(v)-1\}$ such that $\{0,2, \dots, k_1, k_1 + 1, k_1+3, \dots, k_2, k_2+1, k_2+3, \dots, d_{M\oplus N}(v))\} \subseteq B_v$.
\end{enumerate}
\end{observation}

In the following Lemmas we will derive some structure of basic canonical paths, which will be useful in proving Lemma \ref{podst}. We summarize these lemmas in Corollary \ref{cor:summary}.

\begin{lemma}
\label{lem:endpoints_type}
Let $\cal S$ be a basic path, such that its endpoints are distinct. Let $v,u$ be the enpoints of $\cal S$. Then there is no $k\in \{1,2,\dots, d_{\cal S}(v)-2\}$ such that $\{k, k+1\} \subseteq B(v)$.
\end{lemma}
\begin{proof}
The proof will be by contradiction. Assume that there is a $k\in \{1,2,\dots, d_{\cal S}(v)-2\}$ such that $v$ allows $\{k,k+1\}$. Therefore by Observation \ref{obs:degrees_canonical_path} $\{0,2,\dots, k-2, k, k+1, k+3, \dots, d_{\cal S}(v)-2, d_{\cal S}(v)\} \subseteq B(v)$. Let $m\in \{0,1,\dots,d_{\cal S}(v)-1\}$ be such that $\{m,m+1\}\subseteq B(u)$. We will now construct a subset $\S'$ of $\S$ which is a canonical path and such that $w(\S') \geq \min(w(\S),0)$. Let us consider three cases:
\begin{enumerate}
	\item $m = 0$. For any cycle $C$ in $\S$, $\S \setminus C$ is a canonical path. So if $\S$ contains a non-positive cycle $C$ $\S$ is not a basic path. Otherwise $\forall C\quad w(C) > 0$. If there is a cycle incident to $v$ and not incident to $u$ then it is a canonical path with positive weight. If not then there must be a cycle $C$ incident to $u$ and $v$ (as $v$ is incident to at least one cycle). We split $C$ into two paths connecting $u$ and $v$ and we remove the one with smaller weight and the meta-path connecting $u$ and $v$ (from definition of canonical path). We decreased degree of both endpoints by $2$ so it is a canonical path and weight of all cycles and remaining part of $C$ is positive
	\item $m = d_{\S}(v) - 1$. If there is a positive cycle $C\in \S$ then it is a canonical path. Otherwise if there is a cycle $C$ incident only to $v$ then $\S \setminus C$ is a canonical path and $w(\S \setminus C) \geq w(\S)$. Finally if there is no such cycle we take a cycle $C$ incident to $u$ and $v$ and we split it into two paths. The path with greater weight with meta-path connecting $u$ and $v$ forms a canonical path. As we removed some cycles, each of negative weight, and one part of $C$, which also has negative weight, it follows that resulting canonical path has weight greater than $w(\S)$.
	\item $0 < m < d_{\S}(v) - 1$. We take any cycle $C$. It is a canonical path, so if it has positive weight it contradicts the assumption. Otherwise $\S \setminus C$ is also a canonical path and contradicts the assumption.
\end{enumerate}
\end{proof}

Lemma \ref{lem:endpoints_type} shows that if a canonical path $\S$ has distinct endpoints then for its endpoint $v$ either $B(v) \cap \{0,\dots,d_{\S}(v)\} = \{0,1,3,\dots, d_{\S}(v) - 2, d_{\S}(v)\}$ or $B(v) \cap \{0,\dots,d_{\S}(v)\} = \{0,2,4,\dots, d_{\S}(v) - 3, d_{\S}(v) - 1, d_{\S}(v)\}$. The first case happens when $v$ is odd and the second case when it is even.

\begin{lemma}
\label{lem:cycle_both_endpoints}
Let $\S$ be a basic path with distinct endpoints. If $\S$ contains a cycle $C$ incident to both endpoints then one of those endpoints is odd and the other is even.
\end{lemma}
\begin{proof}
The proof will be by contradiction. Let us assume that either both endpoints are odd or both are even. In the first case if $w(C) \leq 0$ then $\S \setminus C$ is a canonical path such that $w(\S\setminus C) \geq w(\S)$. Otherwise we split $C$ into two paths connecting endpoints of $\S$. As $C$ has positive weight, one of those paths also has positive weight and it is a canonical path. The case when both endpoints are even is similar.
\end{proof}

\begin{lemma}
\label{lem:cycle_degrees}
Let $\S$ be a basic path with one endpoint (so the meta-path from the definition of a canonical path is empty and $\S$ is a collection of cycles). Let $v$ be the endpoint of $\S$. Then $\S$ is either (a) a single meta-cycle or (b) $\{0,1,3,\dots,d_{\S}(v) - 1, d_{\S}(v)\} \subseteq B(v)$.
\end{lemma}
\begin{proof}
Assume that for some $0<k<\frac{d_{\S}(v)}{2}$ $2k\in B(v)$. Then let $\S'$ be $k$ cycles of $\S$ of greatest weight. If all of these cycles have positive weight then $\S'$ has positive weight. Otherwise all excluded cycles have non-positive weight, so $w(\S') > w(\S)$. 
\end{proof}

\begin{lemma}
\label{lem:cycle_weight}
Let $\S$ be a basic path with distinct endpoints. Let $u,v$ be endpoints of $\S$. If $v$ is even then all cycles incident to $v$ but not to $u$ have non-positive weight. If $v$ is odd then all cycles incident to $v$ but not to $u$ have positive weight.
\end{lemma}
\begin{proof}
Let $v$ be an even endpoint, and $C$ a cycle incident only to $v$. Then if $w(C) > 0$ then $C$ is a canonical path of positive weight, which means that $\S$ is not a basic path.
 
Similarly if $v$ is an odd endpoint, then we can remove any incident cycles of non-positive weight.
\end{proof}

We summarize those lemmas in the following Corollary. We will often implicitly refer to this Corollary in the proof of Lemma \ref{podst}.

\begin{corollary}
\label{cor:summary}
Let $\S$ be a basic canonical path with endpoints $u$ and $v$. Then:
\begin{itemize}
	\item For every vertex $w$ which is not an endpoint of $\S$, $B(w)$ contains $0, 2, \dots, d_S(w)$;
	\item If $u\neq v$ and $u$ is an odd endpoint, then $B(u)$ contains $0, 1, 3, \dots, d_S(u)$. If $u$ is an even endpoint, then $B(u)$ contains $0, 2, \dots, d_S(u)-1, d_S(u)$;
	\item If $u=v$, then either $\S$ is a single meta-cycle and $B(u)$ contains $0,2$ or $B(u)$ contains $0,1,3,\dots, d_S(u)-1, d_S(u)$;
	\item If $u \neq v$ and $u$ is odd, then any cycle incident only to $u$ is positive. If $u$ is even, then any cycle incident only to $u$ is non-positive;
	\item If $u \neq v$ and $\S$ contains a cycle $C$ incident to both $u$ and $v$, then $u$ is odd and $v$ is even.
\end{itemize}
\end{corollary}

\begin{lemma}
\label{lem:cycle_weight_both_endpoints}
Let $\S$ be a basic path with distinct endpoints and let $v$ be its endpoint. 

If $w(\S) \leq 0$ and  $v$ is even, then $v$ is incident to a cycle of $\S$ of non-positive weight. 
Similarly, if $w(\S) > 0$ and $v$ is odd and incident to any cycle of $\S$, then $v$ is incident to a positive cycle of $\S$.
\end{lemma}

\begin{proof}
Let us assume that $\S$ is non-positive and $v$ is even. If there is a cycle in $\S$ incident only to $v$ (but not the other endpoint) then from Lemma \ref{lem:cycle_weight} it has non-positive weight.
Therefore we assume that there is no cycle incident only to $v$ and let $\C$ be a nonempty set of cycles in $\S$ incident to both endpoints (which means that the other endpoint is odd). If $\C$ contains a cycle of non-positive weight, we are done. Otherwise, all cycles in $\S$ have positive weight (as the other endpoint is odd), so the meta-path $\P$ connecting endpoints of $\S$ has non-positive weight. In this case we can take any cycle of $\C$ and split it into two meta-paths $\P_1$ and $\P_2$ between endpoints of $\S$. Assume $w(\P_1) \geq w(\P_2)$. Then we make $\P_1$ the meta-path of $\S$ and $\P_2\cup \P$ a meta-cycle of $\P$, which has non-positive weight.

The other case is similar. 
\end{proof}

Now we will prove Lemma~\ref{podst}.

\technicallemma*

\begin{proof}
To construct a canonical path $\T$ we will consider how $\Q$ and $\R$ interact with each other, that is what common vertices they have. Firstly let us notice that we can assume that $\Q$ and $\R$ do not have a common vertex $v$ that is not an endpoint of any of them. That is because $v$ allows degrees $0,2,4,\dots, d_{Q\cup R}(v)$. Therefore we can create $k := d_{Q\cup R}(v)/2$ new vertices $v_1, v_2, \dots, v_k$ and replace $v$ with a different vertex in each meta-path or meta-cycle containing $v$. Each of these vertices $v_i$ will allow degrees $\{0,2\}$, if it is an endpoint of some alternating path, or $\{0\}$ otherwise. Then any canonical path we will find in the new graph corresponds to some canonical path in the old graph.

The structure of the proof is as follows. First we  prove some auxiliary lemmas. Then we  split the proof into a few cases depending on the structure of $\Q$ and $\R$. If both $\Q$ and $\R$ have two endpoints we use Lemmas \ref{lematR1} and \ref{lematR}. In the second of these lemmas we assume that $\R$ contains at least two edge-disjoint paths between both endpoints of $\R$. 
If $\R$ has only one endpoint we use Lemma \ref{lem:R_one_endpoint}. Finally, if $\Q$ has one endpoint and $\R$ has two endpoints we use Lemma \ref{lem:Q_one_endpoint}.

We say that a path or cycle {\bf goes through} vertex $b$ if two edges of this cycle or path are incident to $b$. 

\begin{lemma} \label{helpbas}
Let $\S \subseteq \R$ be a path with the endpoints $c$ and $d$ such that (i) $w(\S)>0$, (ii) both $c$ and $d$ belongs to $\Q$, (iii) $\S$ does not go through an even endpoint of $Q$. Then every path contained in $\Q$ between $c$ and $d$ that does not go through any even endpoint of $\Q$ has weight at least $w(\S)$ and thus positive.
\end{lemma}
\begin{proof} Otherwise, we could replace such path with $\S$ and obtain a canonical path of greater weight than $\Q$.
\end{proof}

\begin{lemma} \label{help}
Let $\S \subseteq \R$ be a path with endpoints $c$ and $d$ of positive weight such that both $c$  and $d$ lie on $\Q$.  

 Then, the existence in the graph of any of the listed below implies the existence of a canonical path $\T$ w.r.t. $M$ such that $w(\T) > w(\Q)$:
\begin{enumerate}
\item $\S$ does not go through any odd endpoint of $\Q$ and there exists a path with endpoints $c$ and $d$ contained in $\Q$ that does not go through any  endpoint of $\Q$;
\item $\Q$ has two odd endpoints $a$ and $b$ and $\S$ either goes through $a$ and $b$ or $\Q$ contains a path with endpoints $c$ and $d$ that goes through $a$ and $b$.

\end{enumerate}
\end{lemma}

\begin{proof} In the first case there exists a path $\Q'$ with endpoints $c$ and $d$ contained in $\Q$ that does not go through any endpoint of $\Q$. Then by Lemma \ref{helpbas} $\Q'$ has positive weight and $\Q' \cup \S$ forms a positive cycle that goes only through even vertices and hence is a canonical path w.r.t $M$.

Suppose now that both $a$ and $b$ are odd. Thus $\Q$ contains exactly one path connecting $a$ and $b$.

Assume also that $\Q$ 
 contains a path $\Q'$ with endpoints $c$ and $d$ that goes through $a$ and $b$.  One endpoint of $\Q'$, say $c$  must lie on a cycle $\C_1$ of $\Q$ incident to $a$ and the other - $d$ on a cycle $\C_2$ incident to $\Q$. This means that we can extract from $\C_1$ and $\C_2$ positive  weight paths $\P_1=\P(a,c)$ and $\P_2=\P(b,d)$.  Then  $\S \cup \P_1 \cup \P_2$ is a positive canonical path w.r.t. $M$. Let us notice that this holds regardless if $\S$ goes through $a$ or $b$ or even both of them.

Suppose now that $\S$  contains a path $\P(a,b)$. Therefore $\S$ consists of paths: $\P_1=\P(c,a), \P_0=\P(a,b)$ and $\P_2=\P(b,d)$. If $w(\P_0)>0$, we are done. Otherwise, $w(\P_1) + w(\P_2) > 0$.  By arguments analogous to those used in the proof of Lemma \ref{helpbas} this means that $w(\Q)$ contains two edge-disjoint paths $P'_1=\P(a,c)$ and $\P'_2=\P(d,b)$ such that $w(P'_1) + w(P'_2)\geq w(\P_1) + w(\P_2) > 0$. Then $P'_1 \cup P'_2 \cup \S$ forms a positive canonical path w.r.t $M$. \end{proof}

We denote endpoints of $\Q$ as $a$ and $b$ and endpoints of $\R$ as $c$ and $d$.

\begin{lemma}\label{cykl}
Let $\C \subset \R$ be a cycle with positive weight  that  contains at least one of the endpoints of $\Q$. Then there exists a canonical path of weight greater than $\Q$.
\end{lemma}
\begin{proof} 

\noindent {\bf Case:  $\C$ contains no odd vertex.} $\C$ forms then a canonical path.

\noindent {\bf Case:  $\C$ contains at least  two odd vertices.} Suppose that $\C$ contains $k$ odd vertices. We then split $\C$ into $k$ paths with odd endpoints.  At least one of these paths must have positive weight and forms a canonical path with positive weight.

\noindent {\bf Case:  $\C$ contains exactly one odd vertex $c$ that  belongs to $\R \setminus \Q$.} $\C$ must contain at least one even endpoint  of $\Q$.  We split $\C$ into two paths or three paths depending on whether $\C$ contains one or two even endpoints of $\Q$. We choose the path $\S$ with positive weight. If the endpoints of $\S$ are even endpoints of $\Q$, we are done - by Lemma \ref{help}. Otherwise one of the endpoints of $\S$ is $c$ and the other an even endpoint of $\Q$, let us call it $b$. By Lemma \ref{lem:cycle_weight_both_endpoints} $\Q$ contains a cycle $\C'$ going through $b$ that has non-positive weight. Also, if $\Q$ has two even endpoints $a$ and $b$, then $\C'$ does not go through $a$. Then $\Q \cup \S \setminus \C'$
forms a canonical path with the endpoints $a$ and $c$ and weight greater than that of $\Q$.

\noindent {\bf Case:  $\C$ contains exactly one odd vertex $a$ that  belongs to $\Q$.}
 If $\C$ does not contain a vertex that  is odd w.r.t. $Q$, we can see that $\Q \cup \C$ is of the same uniform type as $\Q$ and has bigger weight.  Assume then that $\C$ contains a vertex that is odd w.r.t. $\Q$.  Let us note that $\C$ cannot contain two vertices that are odd w.r.t. $\Q$ because by Lemma \ref{lem:cycle_both_endpoints} a basic canonical path with two endpoints does not contain a cycle that goes through both endpoints if both of them are odd or both of them are even.

Let us consider first the case when $a=c$ and $a$  is odd w.r.t. $\Q$. We remove from $\Q$ a path between $a$ and $b$ of minimum weight and each cycle incident to $b$ and not going through $a$ - the remaining part of $\Q$ has positive weight or smaller than that of $\Q$. It is so because each cycle contained in $\Q$ going through $a$ and not $b$ has positive weight, each cycle going through $b$ and not $a$ has non-positive weight and either each path between $a$ and $b$ has positive weight or at least one of them has non-positive weight. To thus modified $\Q$ we add $\C$ and obtain a canonical path $\Q'$ with one endpoint $a$ such that $deg_{\Q'}(a)=deg_{\Q}(a) +1$.

Now we assume that $\C$ contains a vertex $d \neq a$ that is odd w.r.t. $\Q$. If $\C$ goes through an even endpoint $b$ of $\Q$, we proceed as follows. By Lemma \ref{lem:cycle_weight_both_endpoints} $\Q$ contains a cycle $\C'$ with non-positive weight going through $b$. Thus $\Q \setminus \C' \cup \C$ forms a canonical path with the endpoints $a$ and $b$ and weight greater than $w(\Q)$. Next we examine the case when $\C$ does not go through any endpoint of $\Q$ different from $a$.

If $\Q$ contains a cycle $\C'$  with non-positive weight going through both $b$ and $d$, where $b$ is even then there exists a path $\P \subset \Q$ between $d$ and $b$ of non-positive weight and we build $\Q'= \Q \cup \C \setminus \P$, which is a canonical path with the endpoints $a$ and $d$ and weight greater than $w(\Q)$. 

Otherwise, we build $\Q'$ as follows - we extract from $\Q$ a path $\S$ between $a$ and $d$ - note that $w(\S)>0$ by Lemma \ref{helpbas} as $\C$ contains a path between $a$ and $d$ of positive weight. Next we add every cycle contained in $\Q$ incident to $a$ but not the one containing $\S$ - each such cycle has positive weight. $\Q'$ also contains $\C$. The weight of $\Q'$ is clearly positive. It is also a canonical path with
the endpoints $a$ and $d$ because the degree of $a$ in $\Q'$ is odd and  $deg_{\Q'}(d)=deg_{\Q}(d) +1$. To see that the degree of $d$ in $\Q'$ is as claimed let us notice that $d$ does not belong to any cycle contained in $\Q$ that goes through $b$  and with non-positive weight, which means that $d$ either lies on a path between $a$ and $b$ or on a cycle incident to $a$. Also, there cannot exist two edge-disjoint paths between $a$ and $b$ going through $d$ because then they would form two edge-disjoint cycles - one going through $a$ and $d$ and the other through $b$ and $d$. If the cycle going through $b$ and $d$ has positive weight, it forms a canonical path because it does not go through any odd vertex. \end{proof}

\begin{lemma} \label{L1}
Let $\C \subset \R$ be a cycle with positive weight incident to $c$ and $c \in \Q$. Then there exists a canonical path $\T$ w.r.t. $M$ such that $w(\T) > w(\Q)$.
\end{lemma}
\begin{proof}
The only case that requires  explanation is when $\C$ does not contain any endpoint of $\Q$. Other cases are covered by Lemma~\ref{cykl} above.
Then $\C$ itself forms a canonical path because it does not go through any odd vertex.
\end{proof}

\begin{lemma} \label{stopien}
Suppose that an endpoint $c$ of $\R$ belongs to $\Q$. Then $c$ is either even w.r.t. $\Q$ or
$deg_{\Q \cup \R}(c) = deg_{\Q}(c)+1$.
\end{lemma}
\begin{proof} By Lemma \ref{cykl}, $\R$ does not contain a positive cycle incident to $c$. By Lemma \ref{lem:cycle_weight_both_endpoints} if an endpoint $c$ of $\R$ is odd and contains a cycle incident to it, then it contains a positive cycle. This shows that indeed $c$ is either even w.r.t. $\Q$ or
$deg_{\Q \cup \R}(c) = deg_{\Q}(c)+1$. 
\end{proof}

\begin{lemma}\label{lematZ}
Let $Z \subseteq \R$ consist of a path between $c$ and $b$ and  cycles incident to $c$  and be such that it does not go through any even endpoint of $\Q$. Also, $w(Z)>0$, $b$ is even and $c$ is fine in $\Q \cup Z$. 
If $Z$ goes through $d$, then $d$ is even.

Then there exists a canonical path w.r.t. $M$ with weight greater than that of $\Q$.
\end{lemma}
\begin{proof} By Lemma \ref{lem:cycle_weight_both_endpoints} $\Q$ contains a cycle $\C$ incident to $b$ of non-positive weight. 


Suppose first that $c$ belongs to $\C$.  It means that $c \in \Q$ and thus  by Lemma \ref{stopien}, $c$ is either even w.r.t. $\Q$ or
$deg_{\Q \cup Z}(c)= deg_{\Q \cup \R}(c) = deg_{\Q}(c)+1$.
We extract from $\C$  a path $\P$ with the endpoints $b$ and $c$ and non-positive weight.  We construct $\Q'=\Q \cup Z \setminus \P$. $\Q'$ is a canonical path with the endpoints $a$ and $b$ because the degrees of $c$ in $\Q'$ and in $\Q$ have the same parity
 and the degree of $b$ is the same in $\Q'$ as in $\Q$. If $c$ does not belong to $\C$, we construct $\Q'=\Q \cup Z \setminus \C$. $\Q'$ is a canonical path with the endpoints $a$ and $c$. In both cases  $w(\Q') > w(\Q)$. 
 \end{proof}

\begin{lemma}\label{lematR1} 
If $\R$ contains a path $\R_{max}$ between $c$ and $d$ and no two edge-disjoint paths between $c$ and $d$, then there exists a canonical path of weight greater than $\Q$.
\end{lemma}
\begin{proof}
The general approach in this proof is the following. We start by considering a set $Z$ consisting of a path $R_{max}$ and every cycle $\C \subset \R$ that does not go through any endpoint of $\Q$. By Lemmas \ref{cykl} and \ref{L1} the weight of $Z$ is positive because every cycle $\C \subset \R$ that we have not included has non-positive weight. If both $c$ and $d$ are fine in $\Q \cup Z$, we consider an appropriate case. Observe that if $c$ is not an endpoint of $\Q$, then $c$ is not fine in $\Q \cup Z$ iff $c$ is even w.r.t. $\Q$ (and thus also even w.r.t. $M$) and some cycle $\C \subset \R$ goes through $c$ and an endpoint of $\Q$.  This is because the degree of $c$ is odd in $\Q \cup Z$. Next we want to add parts of the non-selected cycles to $Z$ to make $c$ and $d$ fine in $\Q \cup Z$ or show directly that a given case implies that $w(\Q)$ is already positive.


\noindent {\bf Case $1$: (i) $d$ is not fine in $\Q \cup Z$ and some cycle of $\R$  goes through $d$ and an even endpoint $b$ of $\Q$ and (ii) $c$ is fine in $\Q \cup Z$ or no cycle of $\R$  goes through $c$ and an even endpoint of $\Q$.}

We add all cycles incident to $c$ contained in $\R$ to $\Q\cup Z$. As a result $c$ is fine in $Z$.
If $d=b$, then we directly  apply  \ref{lematZ}. Otherwise, we consider any cycle $\C \subset \R$ incident to $d$ that goes through an even endpoint $b$. If $\C$ goes through only one even endpoint of $\Q$ we extract from it a path $\P_1$ between $d$ and $b$ of weight at least $w(\C)/2$. $\P_1 \cup Z$ has positive weight and we may apply Lemma \ref{lematZ} to $Z \cup \P_1$. If $\C$ goes through two even endpoints of $\Q$ -
$a$ and $b$ we partition $\C$ into three paths $\P_1=\P(a,d), \P_2=\P(b,d), \P_3=\P(a,b)$.  If $\P_3$ has positive weight, it means that there exists a cycle $\C_1$ of positive weight that goes only through even vertices - it is formed by $\P_3$ and $\P'=\P(a,b) \subseteq \Q$. 
By Lemma \ref{helpbas} $\P'$ has positive weight. If $\P_3$ has non-positive weight, then we may apply Lemma \ref{lematZ} to $Z \cup \P_1$ or
$Z \cup \P_2$.

Let us note that the above arguments hold even if $a=c$. The case $c=b$ does not happen as $\R$ does not contain two edge disjoint paths between 
$c$ and $d$.

\noindent {\bf Case $2$: each of the endpoints of $\R$ is not fine in $\Q \cup Z$ or coincides with one of the endpoints of $\Q$. }

Suppose first that no endpoint of $\R$ coincides with any endpoint of $\Q$.
There exists then a cycle $\C_1 \subset \R$ incident to $c$ that goes through $a$ and  a cycle $\C_2 \subset \R$ incident to $d$ that goes through $b$. We split $\C_1$ into two paths between $a$ and $c$ and choose the one with greater weight - let us call it $\P_1$. Similarly, we split $\C_2$ into two paths between $b$ and $d$ and choose the one with greater weight and call it $\P_2$.

We note that the path $\S=\R_{max} \cup \P_1 \cup \P_2$ between $a$ and $b$ has positive weight. It follows from the fact that each cycle of $\R$ incident to $c$ or $d$ has non-positive weight - recall that each endpoint of $\R$ is even w.r.t. $\Q$.

If both $a$ and $b$ are even or both of them are odd, we are done, as either $\S$ forms a canonical path w.r.t $M$ or by Lemma \ref{help} its existence implies the existence of a positive cycle going only through even vertices. The case when $a$ is odd and $b$ is even is already covered in the preceding case. Let us note that if $R_{max}$ goes through some even endpoint(s) of $\Q$, we may also need to split $\S$.

If one or two endpoints of $\R$ coincide with the endpoints of $\Q$,  then the task of obtaining a path $\S$ between $a$ and $b$ of positive weight is even easier as we do not have to add parts of some cycles of $\R$ incident to respective endpoint(s) of $\R$.

\noindent {\bf Case $3$: (i) $d$ is  fine in $\Q \cup Z$ and lies on a cycle $\C$ of $\Q$  that goes through $d$ and an even endpoint $b$ of $\Q$ but not through $c$ and such that $w(\C) \leq 0$ and (ii) $c$ is fine in $\Q \cup Z$ or no cycle of $\R$  goes through $c$ and an even endpoint of $\Q$.}

We add all cycles incident to $c$ contained in $\R$ to $Z$. As a result $c$ is fine in $\Q\cup Z$ and $Z$ still has positive weight. 

If $\R_{max}$ goes through exactly one even endpoint of $\Q$ we split $Z$ into two parts, choose the one with greater weight and apply Lemma \ref{lematZ} to it.  If  $\R_{max}$ goes through two even endpoints of $\Q$, we split it into three parts and either apply Lemma \ref{lematZ} to one of the parts or, if there exists a path $\P(a,b) \subseteq \R_{max}$ with positive weight, we apply Lemma \ref{helpbas} and obtain a positive cycle that goes only through even vertices.

Otherwise, we extract from $\C$ a path $\P'=\P(b,d)$ of non-positive weight and then $Z \setminus \P'$ forms a canonical path of weight greater than $w(\Q)$.

\noindent {\bf Case $4$: $c \notin \Q$, both  $c$ and $d$ are fine in $\Q \cup Z$.}

Note that $d \in \Q$. Otherwise $Z$ would form a canonical path with positive weight or $Z$ would contain odd endpoints of $\Q$ so we could split it into canonical paths. If $Z$ goes through some even endpoint(s) of $\Q$, we split it and apply Lemma \ref{lematZ} or Lemma \ref{helpbas}.

If $d$ lies on a cycle $\C \subset \Q$  incident to an odd endpoint $a$ of $\Q$, we extract from $\C$ a path $\P$ of positive weight connecting $a$ and $d$ and then $\P \cup Z$ forms a canonical path of positive weight. The case when $d$ lies on a cycle $\C \subset \Q$  going through an even endpoint $b$ of $\Q$ but not through an odd endpoint of $\Q$ is dealt in case $3$.

If $d$ lies on a path $\P$ of $\Q$ between $a$ and $b$ and $\Q$ does not contain two edge-disjoint paths between $a$ and $b$, we split $\Q$ with $d$ into two parts and take either one part of $\Q$ and $Z$ or the other and $Z$ and obtain a canonical path of weight greater than $\Q$ ($d$ is fine by Lemma \ref{stopien}).
If $d$ lies on a path $\P$ of $\Q$ between $a$ and $b$ and $\Q$  contains two edge-disjoint paths between $a$ and $b$, we proceed as follows.
We know that in this case one of the endpoints of $\Q$ is odd and the other even. Suppose that $a$ is odd.
We consider $\P$ and another path $\P_1=\P(a,b) \subseteq \Q$ edge-disjoint with $\P$. $\P \cup \P_1$ either forms a cycle that goes through
$a,b,d$ or contains a cycle that goes through $b$ and $d$ but not $a$. In either case this cycle has positive weight - otherwise we could apply
the previous case to it. If we have a cycle of positive weight that does not go through $a$ - it forms a canonical path, because it is a cycle that does not go through any odd vertex. If we have a cycle $\C \subseteq \Q$ that goes through $a,b,d$, we split it into two paths connecting 
$a$ and $d$ and choose the one with greater weight. Let us call it $\S$. The path $\S$ has positive weight and $Z \cup \S$ forms a canonical path with the endpoints $a$ and $c$ of positive weight.

\noindent {\bf Case $5$: $c \notin \Q$, $c$ is fine in $\Q \cup Z$ and $d$ is not.}

There exists then a cycle $\C \subset \R$ incident to $d$ that contains some endpoint of $\Q$. Let us note that we may assume that $\C$ does not contain any even endpoints of $\Q$ - such cases have already been considered.
Depending on which odd endpoints are contained in $\C$, we are able to extract
from $\C$ either (i) a path $\P$ between $a$ and $d$ such that $w(\P) \geq w(\C)/2$ or (ii) a path between $b$ and $d$ such that $w(\P) \geq w(\C)/2$ or (iii) a path $\P$ between $a$ and $b$ of positive weight.

In the first two cases we construct $\Q_1= Z \cup \P$. Note that $d$ is even in $\Q_1$ and $w(\Q_1)>0$. $d$ is also fine in $\Q_1$ as $deg_{\Q_1}(d) < deg_{\Q}(d)$. Also $c$ is fine in $\Q_1$ as well as $d$ is fine in $\Q\cup \Q_1$ - the degrees of $c$ are the same in $\Q_1$ and $\Q \cup \Q_1$.  Thus $\Q_1$ forms a positive weight canonical path with the endpoints either $a$ and $c$ or $b$ and $c$.
In the last case $\P$ itself forms a canonical path.

\noindent {\bf Case $6$: $a$ and $b$ are odd, a cycle $\C \subset \R$ incident to $c$ goes through $a$ and $b$.} 

It means that $\R$ does not contain any cycle incident to $d$ that goes through $a$ or $b$ and thus that $d$ is fine in $\Q \cup Z$. Also we may assume that $d \in \Q$ - the other case is already covered above. We  split $\C$ into three meta-paths $\P_1=\P(a,b), \P_2=\P(a,c), \P_3=\P(b,c)$.

We observe that every cycle contained in $\R$ has non-positive weight. Therefore $w(R_{max})>0$ because $w(\R)>0$.
 
 We will show that $\C \cup \R_{max}$ contains two  paths $\S_1=\P(a,d)$ and $\S_2=\P(b,d)$, each of which has positive weight. We know that $w(\C \cup \R_{max})>0$. If $w(\P_3) <0$, then $\S_1= \P_1 \cup \P_2 \cup \R_{max}$ has positive weight.
Otherwise $\S_1 = \P_3 \cup \R_{max}$ has positive weight. Similarly, if $w(\P_2) <0$, then $\S_2= \P_1 \cup \P_3 \cup \R_{max}$ has positive weight
and otherwise $\S_2= \P_2 \cup \R_{max}$ has positive weight. Using Lemma \ref{helpbas}, we know that any  path $\P \subset \Q$ between $a$ and $d$ or $b$ and $d$ has positive weight. 

Let $\Q'$ be the path contained in $\Q$ between $a$ and $b$. If it goes through $d$ then it has positive weight. Then $\Q$ also has positive weight as every cycle of $\Q$ is incident to odd enpoint and thus has positive weight.
Otherwise let $C$ be the cycle that contains $d$ and let us assume it is incident to $a$. Then we split $C$ into two paths $C_1$ and $C_2$ between $a$ and $d$. Both $C_1$ and $C_2\cup \Q'$ have positive weight by Lemma \ref{helpbas}, so $\Q$ also has positive weight.

Let us observe that $d$ cannot coincide with either $a$ or $b$ as $\R$ contains only one path between $c$ and $d$. If $c$ coincides with either $a$ or $b$ the arguments above hold.

\noindent {\bf Case $7$: $a$ and $b$ are odd, a cycle $\C_1 \subset \R$ incident to $c$ goes through $a$ and a cycle $\C_2 \subset \R$ incident to $c$ goes through $b$.}

Again, we may assume that $d \in \Q$. We again observe that every cycle contained in $\R$ has non-positive weight. Therefore $w(R_{max})>0$ because $w(\R)>0$.
 
We show that $w(\Q)>0$. To this end it suffices to show that the path $\P(a,b) \subseteq \Q$ has positive weight.
We extract from $\C_1$ and $\C_2$  paths $\P_1$ and $\P_2$, correspondingly between $a$ and $c$ and $b$ and $d$ such that $w(\P_1) \geq w(\C_1)/2$ and
$w(\P_2) \geq w(\C_2)/2$. It means that $w(\R_{max} \cup \P_1 \cup \P_2)>0$. This in turn means by Lemma \ref{helpbas} that that the path $\P(a,b)$ contained in $\Q$ has positive weight.

Similarly as in the case above $d$ cannot coincide with either $a$ or $b$ and if $c$ coincides with either $a$ or $b$ the arguments above hold.

\noindent {\bf Case $8$: $d$ is fine in $\Q \cup Z$, $c$ is not and $c \notin \Q$.} 

It means that there exists $\C \subset \R$ incident to $c$ that goes through exactly one odd endpoint of $\Q$, say $a$, and $d \in \Q$ - other cases are dealt with above. Also, we may assume that $d$ does not lie on a non-positive cycle $\C' \subseteq \R$ that goes through $b$ (that is case $3$).

We proceed as follows. We extract from $\C$ a path $\P_1$ with the endpoints $a$ and $c$ such that $w(\P_1) \geq w(\C)/2$.
$\Q_2$ consists of $Z$ and a path $\P \subseteq \Q$ between $a$ and $d$. $w(\P)>0$ because $w(\P_1 \cup \R_{max}) >0$. Therefore $\Q_2$ has positive weight and is a canonical path with the endpoints $a$ and $c$.

We are left with the following case.

\noindent {\bf Case $9$:  (i) $c,d\in \Q$, $d$ is fine in $\Q \cup Z$}

Let us observe that $b$ is not contained in any cycle $C$ of $R$ - the other case is dealt with above. 

If $\R_{max}$ goes through some even endpoint(s) of $\Q$, then we may split $R$ and apply Lemma \ref{lematZ} or Lemma \ref{helpbas}.
If $\R_{max}$ goes through two odd endpoints of $\Q$, then we may apply Lemma \ref{help}. Thus we may assume that $R_{max}$ goes through at most one endpoint of $\Q$ and if it does, it is through an odd endpoint of $\Q$.

Also, we may assume that either every path connecting $c$ and $d$ contained in $\Q$ goes through some endpoint of $\Q$ or that $\R_{max}$ goes through an odd endpoint of $\Q$ - otherwise we can apply Lemma \ref{helpbas} and obtain a positive cycle going solely through even vertices.

For any endpoint $v$ of $\R$ it holds that if for some edge-set $\Q'$ we have that $deg_{\Q'}(v)= deg_{\Q \cup R}(v)-1$, then $v$ is fine in $\Q'$, because by Lemma \ref{stopien}  $v$ is even w.r.t. $\Q$ or $\R$ does not contain any cycle incident to $c$.

\begin{claim}
If $b$ is even and $\Q$ contains a path $\T$ connecting $b$ and $d$ with non-positive weight and such that $\T$ does not go through $c$ or any even endpoint of $\Q$, then $\Q'=\Q \cup \R \setminus \T$ is a canonical path w.r.t $M$ having weight greater than $w(\Q)$.
\end{claim}
\begin{proof}
Obviously, $w(\Q') > w(\Q)$. 
The degree of $c$  is the same in $\Q'$ and $\Q$. The degree of $c$ is the same in $\Q \cup \R$ and $\Q'$, which is fine. The degree of $b$ is the same in $\Q$ and $\Q'$. The parity of the degree of $a$ is the same in $\Q$ and $\Q'$. Also, $deg_{\Q'}(d)= deg_{\Q \cup \R}(d)-1$. It means that $\Q'$ is a canonical path w.r.t. $M$  and has endpoints $a$ and $c$.
\end{proof}

Of course, in the above claim we might replace $c$ with $d$.

\begin{claim} \label{C2}
If $\Q$ contains a path $\T$ connecting $c$ and $d$ with non-positive weight and such that $\T$ does not go through  any even endpoint of $\Q$, then $\Q'=\Q \cup \R \setminus \T$ is a canonical path w.r.t $M$ having weight greater than $w(\Q)$.
\end{claim}
\begin{proof}
$\Q'$ is a canonical path with the endpoints $a$ and $b$ and weight greater than that of $\Q$.
\end{proof}

\begin{claim} \label{C3}
Suppose that $b$ is an even endpoint of $\Q$.
If $\Q$ contains a cycle $\C$ with non-positive weight going through $b$ and $d$ or $b$ and $c$, then it contains a path $\T$ with non-positive weight such that $\Q_1= \Q \cup \R \setminus \T$ is a canonical path and has weight greater than $w(\Q)$.
\end{claim}
\begin{proof}
If $\C$ does not go through $c$, we partition $\C$ into two paths $\T_1$ and $\T_2$, both with the endpoints $c$ and $b$. If $\C$ goes through both $c$ and $d$, then
we partition $\C$ into three paths, with endpoints correspondingly, $b$ and $c$, $b$ and $d$ and $c$ and $d$. For each one of them it holds
that $\Q \cup \R \setminus \T_i$ is a canonical path. Clearly at least one of the paths has non-positive weight.
\end{proof}

Let us suppose then that we cannot apply any of the above claims.
We are left with the following cases:

\begin{enumerate}
\item $c$ and $d$ both lie on one path connecting $a$ and $b$ in $\Q$ and $\R_{max}$ goes through $a$.
\item $c$ lies on a cycle of $\Q$ incident to $a$ and $d$ on a path between $a$ and $b$ in $\Q$.
\item $c$ and $d$ lie on two different cycles of $\Q$ incident to $a$.
\item $c$ and $d$ lie in two different paths between $a$ and $b$ in $\Q$.
\end{enumerate}

Let us show that these are indeed the only remaining cases. By Claim \ref{C3} we may exclude any case, where an endpoint of $\R$ lies on a cycle
going  through an even endpoint of $\Q$ but not through an odd endpoint of $\Q$ as each such cycle has non-positive weight. The case when $c$ lies on a cycle of $\Q$ incident to an odd endpoint $a$ and $d$ on a cycle of $\Q$ incident to an odd endpoint $b$ can be dealt with using Lemma
\ref{help}.

In each of the remaining four  cases we proceed as follows. We remove from $\Q$: $\T$ - a path contained in $\Q$ connecting $d$ and $b$ and also all cycles going through $b$ but not  through $c$ or $d$. We obtain an edge-set $\Q'$ which is a canonical path w.r.t. $M$ with the endpoints $a$ and $c$.  We show that $w(\Q')>0$.

In the first case it is enough to show that a path $\P$ connecting $a$ and $d$ that belongs to $\Q \cap \Q'$ has non-negative weight.
We split $\P$ and $\R_{max}$ into two paths: correspondingly $\P_1=\P(a,c)$ and $\P_2=\P(c,d)$ and $\S_1=\P(c,a)$ and $\S_2=\P(a,d)$. The weight of 
$\P_2$ is positive because $w(\R_{max})>0$ and by Lemma \ref{helpbas}. It holds that $w(\S_1) >0$ or $w(\S_2) >0$. If $w(\S_2)>0$, then $w(\P)>0$ and we are done.
In the other case, $w(\P_1)>0$ (because $w(\S_1)>0$ and by Lemma \ref{helpbas}).  We also already know that $w(\P_2)>0$, which means that $w(\P)>0$. 

In the second case let us note that any path $\T' \subset \Q$ connecting $c$ and $d$ has positive weight by Lemma \ref{helpbas} and the fact that $w(\R_{max})>0$. Let us notice that the part of $\Q'$ that is contained in $\Q$ consists of one such path $\T'$ and some number of cycles incident to $a$, all of which have non-negative weight. Since $\Q' =(\Q' \cap \Q) \cup \R)$, we are done.

In the third case the cycle $\C$ contained in $\Q$ going through $a$ and $d$ has positive weight and if we split it into two paths connecting $a$ and $d$, while building $\Q'$ we can remove that path, whose weight is not bigger. Therefore $\Q' \cap \Q$ consists of one such path contained in $\C$ and some number of cycles contained in $\Q$ and going through $a$. 
\end{proof}

\begin{lemma}\label{lematR} 
If $\R$ contains two edge-disjoint paths between $c$ and $d$, then there exists a canonical path of weight greater than $\Q$.
\end{lemma}

\begin{proof}
Exactly one of the endpoints of $\R$ is odd w.r.t. $\Q$, assume it is $c$. Let us note that if $c \in \Q$, then $c$ is even w.r.t. $M$.

Suppose first that $R$ contains a cycle $\C$ of positive weight. If $\C$ does not contain any odd vertices, $\C$ constitutes a canonical path and we are done. By Lemma \ref{cykl}, if $\C$ contains any of the vertices $\{a,b\}$, we are also done.

Let us notice that $\R$ always contains some cycle $\C$ of positive weight. Any cycle $\C' \subset \R$ going through $c$ and not $d$ is of positive weight. Such cycle $\C'$ for sure does not go through  $a$ or $b$ (by Lemmas \ref{helpbas} and \ref{help}).  Also, if $\C'$  exists, it means that $c \notin \Q$.
 
If such $\C'$ does not exist, then  $\R$ contains two edge-disjoint paths $\R_1, \R_2$ between $c$ and $d$ such that $w(\R_1 \cup \R_2)>0$. Such edge-set  must contain some cycle $\C$ of positive weight. 

The only possibility that  $\C \subset \R$ of positive weight does not imply the existence of a canonical path with positive weight is when $\C$ goes through $c$, $c$ does not belong to $\Q$ ($c$ is odd) and   goes through neither $a$ nor $b$. For the rest of the proof suppose that this is the case.

Suppose now that some path $\S \subset \R$ between $c$ and $d$ contains some endpoint of $\Q$. We consider  the set $Z \subseteq \R$ that consists of edge-disjoint paths $\S_1, \ldots, \S_k, \S$, each with the endpoints $c$ and $d$ and such that either (i) $\R$ does not contain $\C'$ as above and then no path $\S_i$ contains any endpoint of $\Q$ and $k \geq 2$ or (ii) $\R$  contains some $\C'$ as above and then $Z$ contains additionally every such cycle and $k=1$; also $w(Z) >0$. Let us note that such $Z$ always exists.

Suppose that $\S$ contains  exactly one endpoint of $\Q$ - $a$ which is odd or exactly one even endpoint - $b$ and that $Z$ is as in case (i). If $k$ is odd, we consider $\S' \subset \S$ - a path between $c$ and the distinguished endpoint. If $w(\S') \leq 0$, $Z \setminus \S'$ is either a canonical path
with the endpoints $a$ and $c$ with positive weight or we can apply Lemma \ref{lematZ} to it. If $w(\S') >0$, again $\S'$ is either a canonical path
with the endpoints $a$ and $c$ with positive weight (if the distinguished endpoint is an odd endpoint $a$)  or we can apply Lemma \ref{lematZ} to it. If $k$ is even, we proceed in the same way but considering $\S'' \subset \S$ - a path between $d$ and the distinguished endpoint. 
  
If $\S$ contains two odd endpoints of $\Q$ or two even endpoints, we act similarly but split $\S$ into three paths with the endpoints $a$ and $b$, $a$ and $c$, and $b$ and $d$.

Suppose now that no path $\S \subset \R$ between $c$ and $d$ contains any endpoint of $\Q$.

It means that there exists a cycle $\C \subset \R$ that goes through some endpoint $a$ of $\Q$.  It also goes through $d$ and not $c$ and also has non-positive weight. 
We split $C$ either into three paths $P_1=P(a,b), P_2=P(a,d), P_3=P(b,d)$ - if $\C$ goes also through $b$,
or two paths with the endpoints $a$ and $d$.

If $w(P_1) >0$ and both $a$ and $b$ are even or both $a$ and $b$ are odd, we are done - by Lemmas \ref{helpbas} and \ref{help}. Otherwise we are able to extract
from $\C$ a path with one endpoint equal to $d$ and the other either $a$ or $b$ such that $w(P) \geq w(C)/2$ and $P$ does not go through any even endpoint of $Q$.

We construct $\Q_1$. It consists of  every path $\S \subset \R$ with the endpoints $c$ and $d$.
$P$ and each cycle contained in $R$ incident to $c$ but not $d$. Clearly $w(\Q_1) >0$ as in order to obtain $\Q_1$, we have removed from $\R$ at most $w(\C)/2$ which has non-positive weight and possibly some cycles of $\R$ incident to $d$ but not $c$, each one also with non-positive weight.

Note that $d$ is even in $\Q_1$. It is also fine in $\Q_1$ as $deg_{\Q_1}(d) < deg_{\Q}(d)$. Also $c$ is fine in $\Q_1$ as well as $d$ is fine in $\Q\cup \Q_1$ - the degrees of $c$ are the same in $\Q_1$ and $\Q \cup \Q_1$. 

If $P$ ends at an odd endpoint, say $a$, of $Q$ - $Q_1$ forms a canonical path with the endpoints $a$ and $c$.
Otherwise we can treat $\Q_1$ as $Z$ from Lemma \ref{lematZ}.
\end{proof}

\begin{lemma}
\label{lem:shortcut_one_endpoint}
Suppose that $\Q$ has one endpoint $a$ and there is a positive meta-path $S$ between $a$ and $c$. Suppose also that $c$ is incident to $\Q$ and is fine in $\Q \cup S$ and if $S$ contains $d$ then $d$ is even. Then there exists a canonical path of weight greater than $\Q$.
\end{lemma}
\begin{proof}
If $\Q$ contains a non-positive meta-cycle incident to $c$ we split it into two paths and replace lighter of them with $S$. Otherwise $c$ with all cycles of $\Q$ incident to $c$ is a positive canonical path, because not all cycles of $\Q$ are positive and so $a$ is fine.
\end{proof}

\begin{lemma}
\label{lem:one_endpoint_Z}
Suppose that $\Q$ has one endpoint $a$ and $Z$ contains a meta-path $S$ between $a$ and $c$ and possibly some positive cycles incident to $c$, but not containing $d$. Suppose also that $Z$ is positive, $c$ is fine in $\Q \cup Z$ and if $\S$ goes through $d$ then $d$ is even. Then there exists a canonical path of weight greater than $\Q$.
\end{lemma}
\begin{proof}
If any cycle of $Z$ is incident to $a$ then we use Lemma \ref{L1}. If $c\notin \Q$ then $Z$ is a positive canonical path. If $c\in \Q$ and $Z$ contains some cycle we use Lemma \ref{cykl}. Finally if $c\in \Q$ and $Z$ does not contain any cycle we use Lemma \ref{lem:shortcut_one_endpoint}.
\end{proof}

\begin{lemma}
\label{lem:Q_one_endpoint}
Suppose that $\Q$ has one endpoint $a$ and $\R$ has two endpoints $c$ and $d$. Then there exists a canonical path of weight greater than $\Q$.
\end{lemma}
\begin{proof}
If $\Q$ is a single cycle and $1\notin B(a)$ then $\R\cup \Q$ is a canonical path.
Let $R_{max}$ denote path of $\R$ between $c$ and $d$ of maximum weight.

\noindent {\bf Case: $a=c$ and $d\notin \Q$.} 

If $\R$ contains a positive cycle $C$ incident to $c$, but not to $d$, we replace any non-positive cycle of $\Q$ with $C$. If $d$ is odd, then let $\C$ denote cycles of $\R$ incident to $d$, but not to $c$. Each of them has positive weight, so $\C \cup R_{max}$ also has positive weight and is a canonical path. If $d$ is even and there are at least two paths of $\R$ between $c$ and $d$ then we choose from them two heaviest paths and they form a cycle incident to $c$ and $d$ and we replace one of cycles of $\Q$ with it. Finally if $d$ is even and there is exactly one path of $\R$ between $c$ and $d$ then $R_{max}$ with cycles of $\R$ incident to $d$ form a canonical path of positive weight.

\noindent {\bf Case: $a=c$ and $d\in \Q$.}

If there is a positive cycle of $\R$ incident only to one of its endpoints we use Lemma \ref{L1}. Otherwise set of all paths of $\R$ between $c$ and $d$ have positive weight. If $d$ is odd w.r.t. $\Q$ then we use Lemma \ref{lem:shortcut_one_endpoint} by setting $S=R_{max}$. Otherwise we choose two heaviest paths of $\R$ between $c$ and $d$ and they form a cycle $C$ of positive weight. Then we replace any non-positive cycle of $\Q$ with $C$.

\noindent {\bf Case: $a\notin \R$.} 

If there is a cycle of $\Q$ incident to both endpoints of $\R$ then these endpoints are even or not incident to any cycle of $\R$, as otherwise we would use Lemma \ref{L1}. Therefore set of paths between endpoints of $\R$ has positive weight, so also $R_{max}$ has positive weight. Then we use Lemma \ref{help} with $S=R_{max}$. 

Otherwise, if it exists, let $C$ be the cycle of $\Q$ of non-positive weight incident only to one endpoint of $\R$, say $c$. Once again $c$ is even or not incident to any cycle. Let $\R'$ be $\R$ without cycles incident to $c$, but not to $d$ (its weight is greater that $\R$). Let us split $C$ into two paths between $a$ and $c$ and let $P$ be lighter of them. Then $\R' \cup \Q \setminus P$ is a canonical path of weight greater than $\Q$.

Finally suppose that all cycles of $\Q$ incident to any endpoint of $\R$ are positive. Let $\C$ denote these cycles, let $C$ be one of them incident to $c$ and let $P$ be lighter sub-path of $C$ between $a$ and $c$. Then $\R' \cup \C \setminus P$ is a canonical path of positive weight.

\noindent {\bf Case: $a\in \R$}

If $a$ lies on positive cycle of $\R$ incident to only one of its endpoints then we use Lemma \ref{cykl}. Let us consider the case when $a$ lies on some non-positive cycle $C$ of $\R$ incident to only one of its endpoints. Let $c$ be endpoint of $\R$ incident to $C$, which by Corollary \ref{cor:summary} is even, let $P$ be the heavier subpath of $C$ between $a$ and $c$ and let $\C$ be set of cycles of $\R$ incident to $c$. Let $D$ be any cycle of $\Q$ of non-positive weight. If $D$ does not contain $d$ then $\Q \setminus D \cup \R \setminus \C \cup P$ is a canonical path of weight greater than $\Q$. If $D$ contains $d$, then $d$ is even or not incident to any cycle of $\R$ (otherwise we use Lemma \ref{L1}). Let $P'$ be lighter subpath of $D$ between $a$ and $d$. Then $\Q \setminus P' \cup \R \setminus \C \cup P$ is a canonical path of weight greater than $\Q$.

Now we assume that $a$ does not lie on any cycle of $\R$, but it lies on some path of $\R$. If $\R$ contains only one path, then $a$ splits $\R$ into two paths and let $R'$ be heavier of them. If $R'$ is a canonical path we are done. Otherwise let $c$ be the endpoint of $\R'$. If all cycles of $\Q$ incident to $c$ are positive we add them to $\R'$, thus creating positive canonical path. If $c$ is incident to positive cycle, we use Lemma \ref{L1}. Otherwise path of $\R'$ between $c$ and $a$ is positive, which we will denote as $S$. Let $C$ be non-positive cycle of $\Q$ incident to $c$. We split $C$ into two paths between $a$ and $c$ and let $P$ be lighter of them. Then $\Q \setminus P \cup S$ is a canonical path of weight greater than $\Q$.

Finally let us assume that $\R$ has many paths and therefore $c$ is odd and $d$ is even. Firstly let us assume that in $\R$ there is a positive path $S$ between one of its endpoints and $a$, that does not go through the other endpoint. If the endpoint of $S$ is $c$ we use Lemma \ref{lem:one_endpoint_Z}. If the endpoint of $S$ is $d$ let $R_1$ be the path of $\R$ containing $S$ and let $R_{max}$ be the heaviest path of $\R$, unless $R_1$ is heaviest and then let $R_max$ be the second heaviest path. Let $\C$ denote cycles of $\R$ incident to $c$, but not to $d$. We may assume that $R_1\setminus S$ is non-positive, as we would have used previous case, so $S \cup R_{max} \cup \C$ is positive. If $R_{max}$ is not incident to $a$ we use Lemma \ref{lem:one_endpoint_Z} with $Z = c\notin \Q$ $S \cup R_{max}\cup \C$. If $R_{max}$ is incident to $a$, then we split it into $P_1$ between $a$ and $c$ and $P_2$ between $a$ and $d$. If $P_1 \cup \C$ is positive we use Lemma \ref{lem:shortcut_one_endpoint} with $Z=P_1\cup \C$. Otherwise $P_2 \cup S$ is positive so we choose any non-positive cycle of $\Q$ and replace it with $P_2 \cup S$.

In case when all paths of $\R$ incident to $a$ are non-positive, let $\S$ denote those paths that are not incident to $a$ (it might be empty) and let $\C$ be cycles incident to $c$. We can assume that all paths of $\R$ between $c$ and $a$ are non-positive, because otherwise we use Lemma \ref{lem:one_endpoint_Z}. Let us consider $\S \cup \C$, which has positive weight as we only removed non-positive paths and cycles incident to even endpoint. If it is a canonical path we are done. If $c$ is fine and $d$ is not we choose any path between $a$ and $d$ and add it to $\R\setminus\S$ (we still remove only non-positive paths). If $d$ is fine and $c$ is not then let $P$ be a path between $a$ and $c$ and let $R_1$ be path between $c$ and $d$ containing $P$. If $R_1\setminus P$ is positive then $R_1\setminus P$ with maximum path of $\S$ and $\C$ is positive (because either maximum path is positive or we remove only non-positive paths) so we use Lemma \ref{lem:one_endpoint_Z}. Otherwise $\S\cup\C\cup P$ is a positive canonical path. Both $c$ and $d$ are not finee in $\S \cup \C$ only if $\S$ contains odd number of paths and $c\in \Q$. Then we choose a cycle of $\R$ incident to $c$ (either one of $\C$, or if it is empty we form cycle from two heaviest paths between $c$ and $d$) and use Lemma \ref{L1}.
\end{proof}

\begin{lemma} 
\label{lem:R_one_endpoint}
Suppose that $\R$ has one endpoint, denoted by $c$. Let $a$ and $b$ be the endpoints of $\Q$ (if $\Q$ has only one endpoint it will be denoted as $a$). Then there exists a canonical path of weight greater than $\Q$.
\end{lemma}
\begin{proof}
If $\R$ is a single meta-cycle and $1\notin B(c)$ then $\Q\cup \R$ is a canonical path.

We know that $\R$ has positive weight, so there is at least one cycle in $\R$ of positive weight. In such case, if $c$ lies on $\Q$ then by Lemma \ref{L1} we are done.

If both endpoints of $\Q$ are fine in $\R$ then $\R$ is a canonical path with respect to $M$.

Now we assume that one endpoint of $\Q$, say $a$, is not fine in $\R$. $a$ is incident to some number of cycles of $\R$. If any of them, let us call it $C$, has non-positive weight, then we split it into two meta-paths between $a$ and $c$. We remove the lighter of these paths and obtain that way a canonical path, as degree of $c$ is odd and we decreased degree of $a$ by $1$.

If all cycles of $\R$ incident to $a$ have positive weight, we consider any of them and let us call it $C$. We split $C$ into two paths, the same way as before. Let us consider heavier of these paths and call it $P$. Let $\C$ be set of cycles of $\R$ incident to $a$ except the one with $P$. Then $\C \cup P$ is a positive canonical path.

In case when both endpoints of $\Q$ are not fine in $\R$ and incident to $\R$ we proceed similarly. If there is no cycle incident to both $a$ and $b$ and all cycles incident to $a$ are non-positive then we remove all those cycles and proceed with $b$ as before. If all cycles incident to $a$ are positive we form a canonical path from all them except one path between $a$ and $c$ (similarly as above). If there is cycle $C$ incident to both $a$ and $b$ we consider sub-path of $C$ between $a$ and $b$ that does not contain $c$. If it is non-positive we remove it and obtain a positive canonical path. Otherwise we use Lemma \ref{help}.
\end{proof}
\end{proof}
\end{appendix}

\end{document}